\documentclass[12pt]{iopart}
\usepackage{iopams}
\expandafter\let\csname equation*\endcsname\relax
\expandafter\let\csname endequation*\endcsname\relax
\usepackage{graphicx}
\usepackage{booktabs}
\usepackage{amssymb,bm,mathrsfs,bbm,amscd}
\usepackage[tbtags]{amsmath}

\begin{document}

\title{Preliminary Sensitivity Study for a Gravitational Redshift Measurement with China's Lunar Exploration Project}

\author{Cheng-Gang Qin$^{1}$, Tong Liu$^{2,*}$, Xiao-Yi Dai$^{1}$, Peng-Bin Guo$^{2}$, Weisheng Huang$^{1}$, Xiang-Pei Liu$^3$, Yu-Jie Tan$^{1,*}$, and Cheng-Gang Shao$^{1,*}$}

\address{$^{1}$ MOE Key Laboratory of Fundamental Physical Quantities Measurement $\&$ Hubei Key Laboratory of Gravitation and Quantum Physics, PGMF and School of Physics, Huazhong University of Science and Technology, Wuhan 430074, People's Republic of China\\
$^{2}$ Key Laboratory of Space Utilization, Technology and Engineering Center for space Utilization, Chinese Academy of Sciences, Beijing 100094, China\\
$^{3}$ Hefei National Research Center for Physical Sciences at the Microscale and School of Physical Sciences, University of Science and Technology of China, Hefei 230026, China }

\ead{{liutong2021@csu.ac.cn,yjtan@hust.edu.cn,cgshao@hust.edu.cn}}\vspace{10pt}
\begin{indented}
\item[] February 1, 2024
\end{indented}

\begin{abstract}
General relativity (GR) is a highly successful theory that describes gravity as a geometric phenomenon. The gravitational redshift, a classic test of GR, can potentially be violated in alternative gravity theories, and experimental tests on this effect are crucial for our understanding of gravity. In this paper, considering the space-ground clock comparisons with free-space links, we discuss a high-precision Doppler cancellation-based measurement model for testing gravitational redshift. This model can effectively reduce various sources of error and noise, reducing the influences of the first-order Doppler effect, atmospheric delay, Shapiro delay, etc. China's Lunar Exploration Project (CLEP) is proposed to equip the deep-space H maser with a daily stability of $2\times10^{-15}$, which provides an approach for testing gravitational redshift. Based on the simulation, we analyze the space-ground clock comparison experiments of the CLEP experiment, and simulation analysis demonstrates that under ideal condition of high-precision measurement of the onboard H-maser frequency offset and drift, the CLEP experiment may reach the uncertainty of $3.7\times10^{-6}$ after a measurement session of 60 days. Our results demonstrate that if the issue of frequency offset and drift is solved, CLEP missions have a potential of testing the gravitational redshift with high accuracy. This manuscript has been accepted for publication in Classical and Quantum Gravity. DOI 10.1088/1361-6382/ad4ae2
\end{abstract}

\vspace{2pc} \noindent{\it Keywords}: gravitational redshift, fundamental physics, doppler cancellation scheme, deep space clocks

\section{introduction}
General relativity (GR) and other metric theories of gravitation provide a geometric description of gravity interactions \cite{will2014confrontation}. The description of gravitational geometry is based on the Einstein Equivalence Principle (EEP), which assumes that all standard matter couples to the gravitational field in the same way. The EEP is essential in distinguishing metric theories from non-metric theories, as metric theories describe gravity as a geometric phenomenon involving the curvature of spacetime itself. Although EEP has successfully survived all experimental tests so far, there are theoretical motivations to question the validity of the EEP \cite{will2014confrontation}. For example, attempts to unify gravity with the other three fundamental forces such as string theory and quantum gravity theory predict a violation of the EEP \cite{taylor1988dilaton,damour1994string,rubakov2001large,maartens2010brane}. The problem of dark matter and dark energy also suggests that general relativity may be modified in some ways. Therefore, high-precision experimental tests of the EEP are necessary for advancing our understanding of fundamental physics.

The EEP consists of three sub-principles: the universality of free fall (also known as the weak equivalence principle, WEP), local Lorentz invariance (LLI), and local position invariance (LPI) \cite{will2014confrontation}. The WEP states that the motion of an uncharged test body is independent of its internal structure and composition, given an initial velocity and position. The LLI states that the results of any local non-gravitational experiment are independent of the velocity and direction of the motion of reference frame in which the experiment is performed. The LPI states that the results of any local non-gravitational experiment are independent of the position and time of the reference frame in which the experiment is performed.  Precise testing of these sub-principles is crucial for improving our understanding of gravity and distinguishing between different gravitational theories. While advancements in experimental techniques have allowed for high-precision tests of the WEP and LLI \cite{wagner2012torsion,PhysRevLett.100.041101,PhysRevLett.129.121102,PhysRevLett.123.231102,PhysRevLett.118.221102,qin2023testing,sanner2019optical}, the precision of LPI or gravitational redshift tests lags behind. Therefore, it is essential to improve the precision of LPI testing.

For the test of LPI, there are two distinct types of experimental tests. The first type is the traditional gravitational redshift, which involves comparing the frequencies of two identical clocks held at different gravitational potentials. The second type is called a null-redshift test which involves monitoring the variation of the relative frequency of two different kinds of clocks as they move through the gravitational potential together. Long-term atomic transition frequency comparisons have provided stringent limits on violation parameters in null-redshift test experiments \cite{PhysRevLett.109.080801,PhysRevA.87.010102,ashby2018null,PhysRevLett.126.011102}. Gravitational redshift, as a test of LPI, is considered a crucial test of general relativity and alternative gravitation theories. Early successful experiments, such as the Pound-Rebka-Snider experiments conducted in the 1960s and shift measurements of solar spectral lines, tested the validity of gravitational redshift \cite{PhysRevLett.3.439,PhysRevLett.4.337,PhysRev.140.B788}. The Gravity Probe A (GPA) experiment in 1976 compared the frequency difference between an onboard hydrogen maser and a ground-based hydrogen maser, achieving a precision of $1.4\times10^{-4}$ in gravitational redshift testing \cite{PhysRevLett.45.2081}. Recently, space-test experiments based on Galileo satellites achieved an uncertainty of $2.48\times10^{-5}$ \cite{PhysRevLett.121.231101,PhysRevLett.121.231102}, while ground-based experiment with Tokyo Skytree obtained a test precision of $9.1\times10^{-5}$ \cite{takamoto2020test}. Additionally, many clock experiments involving gravitational redshift, with the aid of precise atomic clocks, have been conducted \cite{takano2016geopotential,grotti2018geodesy,bothwell2022resolving,zheng2023lab,zheng2022differential}.

Atomic clocks, based on quantum transitions, provide the best available references for time and frequency, with stability and uncertainty levels reaching $10^{-18}$ and even $10^{-19}$ \cite{hinkley2013atomic,bloom2014optical,PhysRevLett.116.063001,mcgrew2018atomic,PhysRevLett.123.033201,PhysRevApplied.17.034041}. These clocks have extensive applications in fundamental physics tests. Various space missions have been proposed to test general relativity by utilizing atomic clocks in space, including the Earth-orbiting missions like Atomic Clock Ensemble in Space (ACES) \cite{meynadier2018atomic,savalle2019gravitational}, Space Optical Clock (SOC) \cite{bongs2015development}, fundamental physics with a state-of-the-art optical clock in space (FOCOS) \cite{derevianko2022fundamental}, and the China Space Station (CSS) \cite{PhysRevD.108.064031}. These Earth-orbiting missions aim to achieve higher-precision tests of gravitational redshift. In addition, deep-space missions also are proposed to perform the precise measurement of the gravitational redshift, such as the VERITAS mission with Deep Space Atomic Clock (DSAC) \cite{PhysRevD.107.064032}. The space VLBI RadioAstron(RA) mission recently reported a test of gravitational redshift with the uncertainty $3.3\times10^{-4}$ \cite{nunes2023gravitational}. China's Lunar Exploration Project (CLEP) is actively underway for the purpose of lunar explorations and scientific experiments, which also includes proposal of using an H-maser clock with stability of $5\times10^{-13} /\sqrt{\tau}$ (where $\tau$ is the integration time in seconds), and using the microwave links for the time and frequency transfer between the CLEP satellites and the ground stations, such as the Chinese VLBI network \cite{zheng2018technical,zhi2022experiment}. The deep-space CLEP mission with onboard clocks provides a unique opportunity for precise deep-space-mission measurement of gravitational redshift.

In this paper, we explore the potential of CLEP mission to test gravitational redshift, and  discuss a simulation of a test of gravitational redshift through the space-ground clock comparisons in the CLEP. In Sec.\ref{grtest}, we review the theoretical framework of the gravitational redshift test and clock comparison. Sec.\ref{dcf} presents a analysis of ``Doppler Cancellation Scheme" for the clock-comparison experiments of testing gravitational redshift. In Sec.\ref{stgr}, we estimate the accuracy of a potential gravitational redshift with simulated CLEP measurement. Finally, we give the conclusion in Sec.\ref{concl}.

\section{Gravitational redshift and clock comparison}\label{grtest}

According to Einstein's general relativity, the rate of a clock ticks is not constant and it is altered by gravitational fields. On the surface of the Earth, the clocks at higher altitudes tick faster than other identical clocks at lower altitudes, which is known as the famous gravitational redshift. Clock rate comparisons can be achieved by exchanging electromagnetic signals, and gravitational redshift tests can be conducted through clock-comparison experiments using high-precision frequency links.
Let's consider two clocks located at positions $A$ and $B$, where $U_A$ and $U_B$ represent the gravitational potentials at these positions, respectively. Due to the difference in gravitational potentials ($\Delta U = U_B-U_A$), in the presence of gravitational redshift violation, the frequency difference $\delta f$ between the two clocks is given by \cite{will2014confrontation}
\begin{equation}\label{gr1}
  \frac{\delta f}{f}= (1 + \alpha)\frac{\Delta U}{c^2},
\end{equation}
where $c$ is the speed of light and $\alpha$ characterizes the deviation from general relativity, representing the strength of gravitational redshift violation (in general relativity, $\alpha$ equals 0). The magnitude of the gravitational redshift is directly proportional to the difference in gravitational potentials. Therefore, experimental tests of gravitational redshift require maximization of the difference in gravitational potentials or the use of clocks with higher precision for measurement. Near the Earth's surface, the difference in gravitational potential can be expressed as $\Delta U=g\Delta h$, where $g$ is the gravitational acceleration at the surface and $ \Delta h$ is the height difference between the two positions. Ground-based experiments aiming to test gravitational redshift face challenges in achieving large height differences due to building structures and terrain limitations, making it difficult to test gravitational redshift at the $10^{-6}$ level using ground-based clocks.

With the development of aerospace technology, satellite clocks and high-precision time or frequency links have been utilized in space missions to achieve precise clock comparisons between space-based and ground-based clocks. In experiments comparing clocks in space and on the ground, the gravitational potential difference between satellite clocks and ground-based clocks is significant, leading to a more prominent gravitational redshift effect. This advantageous characteristic has been used in the past and offers the potential to improve the precision of gravitational redshift tests and tests of general relativity even further. China has undertaken various lunar exploration projects, such as the Chang'e project and VLBI network, which will deploy high-precision atomic clocks and have the potential to perform tests of gravitational redshift at a high level of accuracy.

In the following sections, we analyze through simulation the potential of tests of gravitational redshift in these deep space exploration missions, focusing on a case within the CLEP experiment that performs the clock comparison between the CLEP satellite and the ground station. Actual deep space missions will face some engineering and technical challenges that will affect the test of gravitational redshift, such as the high-precision measurement of frequency offset and drift of onboard clocks. In this paper, we aim to discuss the relativistic and non-relativistic effects of clock comparison model with the CLEP mission, and explore the potential of CLEP mission to test gravitational redshift under ideal condition of high-precision measurement of onboard-clock frequency offset and drift. Therefore, we do not discuss these challenges.
For the CLEP experiment, we set that the CLEP satellite is in the Distant Retrograde Orbit (DRO) with a resonance ratio of 2:1, the distance from the satellite to the center-of-mass of the Earth is $(2.8\sim4.8)\times 10^{8}$ m, and the distance from the satellite to the center-of-mass of the Moon is $(7\sim10)\times 10^{7}$ m. The DRO has the orbital characteristics of long-term stability and low orbital insertion energy, which ensures the long-term stable operation of the CLEP mission. In the CLEP satellite, the onboard clock is a passive hydrogen maser with stability of $5\times10^{-13} /\sqrt{\tau}$. This onboard hydrogen maser is used as a frequency reference for the one-way signal. For the two-way signal, the onboard electronics or communication system are coherently tied to the uplink signal and thus the ground H-maser. Comparing the experiments in orbit around Earth, the orbital period of CLEP satellite is longer. For a given ground station, the averaged observable time of CLEP satellite can reach several hours, which is an advantage in assessing the stability of atomic clock.

We consider a clock comparison experiment with two clocks in the Solar System Barycentric Celestial Reference System (BCRS), clock $A$ and clock $B$, which compare their frequencies through a light signal. The clock $A$ emits a light signal with a frequency $f_{A}$ at position $x_{A}$, which then is received by clock $B$ at position $x_{B}$, and the received frequency measured at $x_{B}$ is denoted as $f_{B}$. Generally, the difference in frequencies between the two clocks can be characterized by the frequency shift $f_A/f_B$. The general form of the one-way frequency shift between the two clocks is given by \cite{blanchet2001relativistic,PhysRevD.66.024045,qin2019relativistic,jia2023investigation}
\begin{equation}\label{cl1}
  \frac{{f_A}}{{f_B}} = \left(\frac{{d\tau}}{{dt}}\right)_{t_B} \left(\frac{{d\tau}}{{dt}}\right)_{t_A}^{-1} \left(\frac{{dt_B}}{{dt_A}}\right),
\end{equation}
where $\tau_A$ and $\tau_B$ are the proper times of clock $A$ and $B$, respectively, $t_{A}$ represents the coordinate time corresponding to the moment when clock $A$ emits the light signal, and $t_{B}$ represents the coordinate time corresponding to the moment when clock $B$ receives the light signal. The terms ${d\tau_A}/dt_A$ and ${d\tau_A}/dt_B$ in the equation are dependent on the clock's state and mainly include gravitational redshift and second-order Doppler effects, among others. These two terms can be calculated using the invariance of Riemannian space-time intervals. For a clock, it is represented as
\begin{equation}\label{cl2}
 ds^2 = g_{00}c^2 dt^2 + 2g_{0i}c dt dx^i + g_{ij}dx^i dx^j = c^2 d\tau^2,
\end{equation}
where $i$ and $j$ take values 1, 2, and 3. Another term, $dt_A/dt_B$, contains the non-relativistic Doppler effects, which is related to the transmission path of the light signal. For example, in clock experiments on the ground where the clocks are connected by optical fibers, the signal's path is determined by the optical fiber. In space-ground clock comparison experiments, the clocks are compared through microwave or optical links, and the signal's path is determined by the transmission path of the light signal in free space. Clearly, this term also includes first-order Doppler effects, atmospheric delay effects, Shapiro delay effects, temperature effects, etc.

\section{Doppler Cancellation Scheme}\label{dcf}
\subsection{Measurement of Doppler Cancellation Scheme}
\begin{figure}
  \centering
  \includegraphics[width=0.7\textwidth]{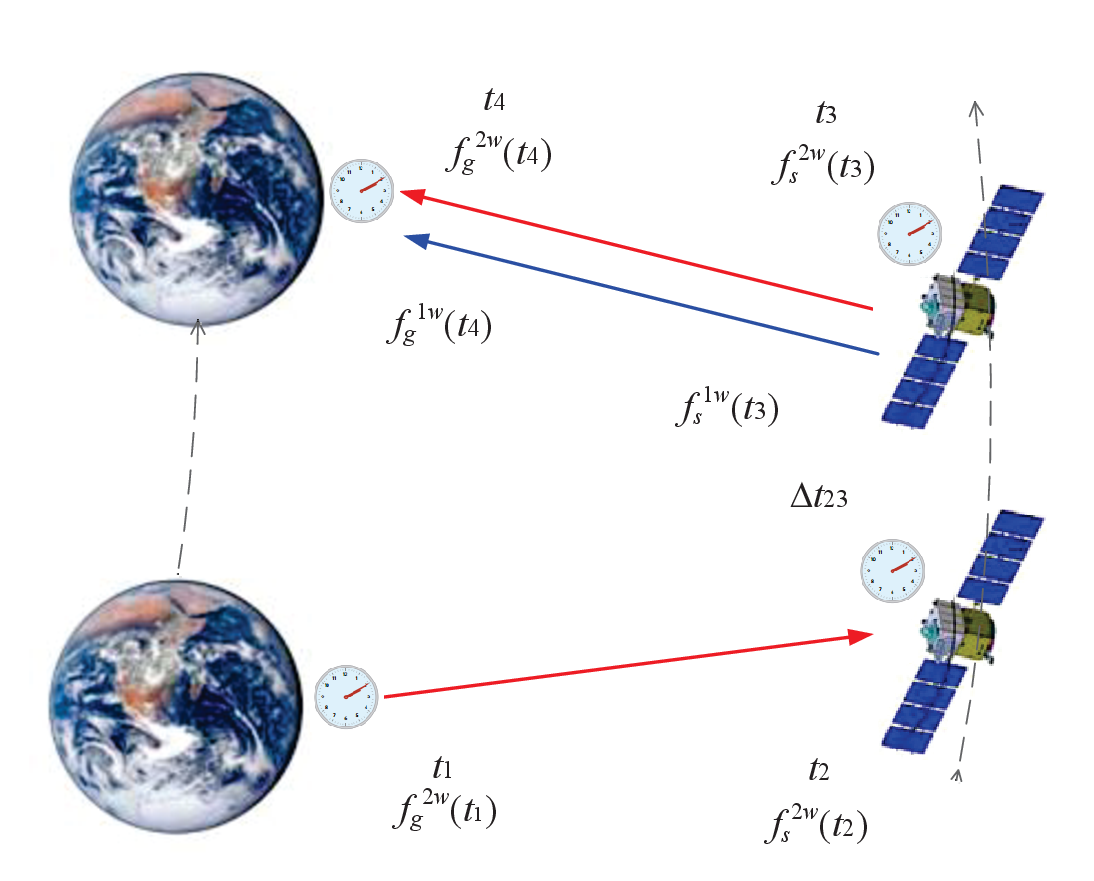}
  \caption{The schematic diagram for clock comparison between the satellite and ground station. The two red signals make up the two-way frequency signal that includes twice contribution of the first-order Doppler shift. The blue line is the one-way frequency signal that contains the first-order Doppler shift. The first-order Doppler shift can be removed by subtracting one-half the two-way Doppler shift from the one-way Doppler shift.   }\label{fig1}
\end{figure}

To achieve a high-precision test of gravitational redshift, the Doppler Cancellation Scheme (DCS) is adopted in the experiments of space-ground clock comparison, which is constructed by one-way and two-way frequency links. DCS technology had been successfully verified by the GPA experiment, which obtained a high precision test of gravitational redshift  \cite{PhysRevLett.45.2081}. As shown in Fig.\ref{fig1}, the space-ground experiment compares the frequencies of clocks in satellite and on the ground station. We firstly consider the two-way frequency transfer, as showed with red lines in Fig.\ref{fig1}. At coordinate time $t_{1}$, the clock on the ground station emits a signal of frequency $f_{g}^{2w}(t_1)$ to the satellite. After being transmitted in free space, this signal is received by the satellite with the frequency $f_{s}^{2w}(t_2)$ at coordinate time $t_2$. This signal gives an uplink from the ground station to the satellite. In this uplink of the two-way frequency transfer, the emission frequency $f_{g}^{2w}(t_1)$ and the reception frequency $f_{s}^{2w}(t_2)$ satisfy the relationship
\begin{equation}\label{cl3}
f_g^{2w}(t_1)d\tau_{g}(t_1) = f_s^{2w}(t_2)d\tau_{s} (t_2),
\end{equation}
where $\tau_{g}(t_1)$ and $\tau_{s}(t_2)$ are the proper times for clocks on the ground and in the satellite, respectively. In the satellite, this signal subsequently is retransmitted to the ground station with retransmission frequency $f^{2w}_{s}(t_3)$ at coordinate time $t_{3}$ and finally received by the ground station with received frequency $f_{g}^{2w}(t_4)$ at coordinate time $t_4$. Similarly, in the downlink of two-way signal, the retransmission frequency $f^{2w}_{s}(t_3)$ and received frequency $f_{g}^{2w}(t_4)$ satisfy
\begin{equation}\label{cl4}
f_{s}^{2w}(t_3)d\tau_{s}(t_3) = f_{g}^{2w}(t_4)d\tau_{g} (t_4),
\end{equation}
where $\tau_{s}(t_3)$ and $\tau_{g}(t_4)$ are the proper times for clocks in the satellite and on the ground, respectively.
The signal forwarding on the satellite is a coherent process, which means that the received signal and the retransmitted signal remain consistent (the delay of the signal forwarding process is $t_{3}-t_{2}$). This process satisfies the relationship $f_{s}^{2w}(t_3)d\tau_s(t_3) = f_s^{2w}(t_2)d\tau_{s}(t_2)$.

Considering frequency transfers in the uplink Eq.(\ref{cl3}) and downlink Eq.(\ref{cl4}), the two-way frequency transfer is obtained. Through two-way frequency transfer and coherent signal forwarding, the signal emitted by the ground station at time $t_1$ returns to the ground station at time $t_4$. In this two-way frequency transfer, the reception frequency at the ground station is given by
\begin{equation}\label{2wf}
f_g^{2w}(t_4) = f_g^{2w}(t_1) \frac{d\tau _g(t_1)}{d\tau _g(t_4)}.
\end{equation}

The one-way frequency transfer is showed with blue line in Fig.\ref{fig1}. This one-way signal is transmitted from satellite at coordinate time $t_3$ to the ground station at coordinate time $t_4$. Similar, the reception frequency in one-way frequency transfer is given by
\begin{equation}\label{1wf}
f_g^{1w}(t_4) = f_s^{1w}(t_3) \frac{d\tau _s(t_3)}{d\tau _g(t_4)}.
\end{equation}

To measure the gravitational redshift, we use the one-way frequency transfer Eq.(\ref{1wf}) and two-way frequency transfer Eq.(\ref{2wf}) to construct a DCS measurement as follows
\begin{equation}\label{dcs1}
\left(\frac{\delta f}{f}\right)_{\text{dcs}} = \frac{f_g^{1w}(t_4)-f_s^{1w}(t_3)}{f_s^{1w}(t_3)}-\frac{f_g^{2w}(t_4)-f_g^{2w}(t_1)}{2f_g^{2w}(t_1)}.
\end{equation}
This equation is an important measurement scheme for gravitational redshift in space missions. In the right-hand side of this equation, the first term represents the one-way frequency shift, which includes the gravitational redshift, Doppler effect, atmospheric delay, etc. The second term represents the two-way frequency shift, which includes twice contributions of the Doppler frequency shift, atmospheric delay, etc. However, in the two-way frequency shift, the gravitational redshift and time dilation effect are canceled out to first order and higher-order residual effects are very small. Through the Doppler cancellation scheme, the influences of the first-order Doppler effect, and atmospheric dealy are greatly suppressed, while the gravitational redshift effect is retained. Therefore, the DCS measurement is applied to the experimental test of the gravitational redshift.

In order to develop Eq.(\ref{dcs1}), it is necessary to calculate the differential relationship between proper time and coordinate time, taking into account the clocks on the ground and the satellite. The theoretical proper times of these clocks are denoted as $\tau_g$ and $\tau_s$ respectively. Generally, there exists a difference between the ideal proper time and the proper time of the signal, which can be expressed as,
\begin{equation}\label{cl5}
  \tilde{\tau}_g = \tau_g + \delta\tau_g^t + \delta\tau_g^e,
\end{equation}
\begin{equation}\label{cl6}
  \tilde{\tau}_s = \tau_s + \delta\tau_s^t + \delta\tau_s^r.
\end{equation}
where, $\delta\tau_g^t$ and $\delta\tau_s^t$ represent the timing errors, which are the differences between the recorded times of the ground station and the satellite and the ideal proper times of their clocks. $\delta\tau_g^e$ and $\delta\tau_s^r$ represent the instrumental delays of the ground station and the satellite (the instrumental delays of signal's emission and reception are different). In the BCRS reference frame $(ct, x^j)$, the evolution of the proper times $\tau_g(t_1)$ and $\tau_s(t_2)$ can be expressed as:
\begin{equation}\label{cl7}
  \frac{{d\tilde{\tau_g}(t_1)}}{{dt_1}} = \left(1 + \frac{{d\delta\tau_g^t}}{{dt_1}} + \frac{{d\delta\tau_g^e}}{{dt_1}}\right) \frac{{d\tau_g}}{{dt_1}},
\end{equation}
\begin{equation}\label{cl8}
  \frac{{d\tilde{\tau_s}(t_2)}}{{dt_2}} = \left(1 + \frac{{d\delta\tau_s^t}}{{dt_2}} + \frac{{d\delta\tau_s^r}}{{dt_2}}\right) \frac{{d\tau_s}}{{dt_2}}.
\end{equation}
These two equations neglect the high-order terms $(d\delta\tau^{t/e}/dt)(1-d\tau/dt)$, which are negligible for the accuracy of current clocks. The differential relationships for coordinate times $t_3$ and $t_4$ can be obtained by analogous expressions.

To compute the DCS measurement equation, it is necessary to compute the one-way frequency transfer and the two-way frequency transfer. According to Eqs.(\ref{2wf})-(\ref{cl8}), the one-way frequency shift can be expressed as
\begin{equation}\label{cl9}
  \frac{{f_g^{1w}(t_4) - f_{s}^{1w}(t_3)}}{{f_{s}^{1w}(t_3)}} = \left(1 + \frac{{d\delta\tau_s^t}}{{dt_3}} + \frac{{d\delta\tau_s^e}}{{dt_3}} - \frac{{d\delta\tau_g^t}}{{dt_4}} - \frac{{d\delta\tau_g^r}}{{dt_4}}\right) \left(\frac{{d\tau_s}}{{dt}}\right)_{t_3} \left(\frac{{d\tau_g}}{{dt}}\right)_{t_4}^{-1} \frac{{dt_3}}{{dt_4}} - 1.
\end{equation}
Similarly, the two-way frequency shift can be expressed as
\begin{equation}\label{cl10}
  \frac{{f_g^{2w}(t_4) - f_g^{2w}(t_1)}}{{f_g^{2w}(t_1)}} = \left(1 + \frac{{d\delta\tau_g^t}}{{dt_1}} + \frac{{d\delta\tau_g^e}}{{dt_1}} - \frac{{d\delta\tau_g^t}}{{dt_4}} - \frac{{d\delta\tau_g^r}}{{dt_4}}\right) \left(\frac{{d\tau_g}}{{dt}}\right)_{t_1} \left(\frac{{d\tau_g}}{{dt}}\right)_{t_4}^{-1} \frac{{dt_1}}{{dt_4}} - 1.
\end{equation}
Based on the two-way frequency shift (\ref{cl9}) and the one-way frequency shift (\ref{cl10}), we can obtain the DCS measurement equation
\begin{equation}
\begin{split}\label{cl11}
  \left(\frac{{\delta f}}{{f}}\right)_{\text{dcs}} &=  \left(1 + \frac{{d\delta\tau_s^t}}{{dt_3}} + \frac{{d\delta\tau_s^e}}{{dt_3}} - \frac{{d\delta\tau_g^t}}{{dt_4}} - \frac{{d\delta\tau_g^r}}{{dt_4}}\right) \left(\frac{{d\tau_s}}{{dt}}\right)_{t_3} \left(\frac{{d\tau_g}}{{dt}}\right)_{t_4}^{-1} \frac{{dt_3}}{{dt_4}} \\
&- \frac{1}{2} \left(1 + \frac{{d\delta\tau_g^t}}{{dt_1}} + \frac{{d\delta\tau_g^e}}{{dt_1}} - \frac{{d\delta\tau_g^t}}{{dt_4}} - \frac{{d\delta\tau_g^r}}{{dt_4}}\right) \left(\frac{{d\tau_g}}{{dt}}\right)_{t_1} \left(\frac{{d\tau_g}}{{dt}}\right)_{t_4}^{-1} \frac{{dt_1}}{{dt_4}}-\frac{1}{2}.
\end{split}
\end{equation}
Eqs.(\ref{cl9}), (\ref{cl10}), and (\ref{cl11}) provide the one-way frequency measurement, two-way frequency measurement, and DCS measurement, respectively. These measurement equations can be applied to the analysis of gravitational redshift tests in various space missions. In one-way frequency measurement and two-way frequency measurement, the first-order Doppler effect can reach the order of $10^{-6}-10^{-5}$ in the CLEP mission (Table.\ref{table1}). Some space missions require the evaluation of clock frequency shifts to the level of $10^{-16}$ or better, and a priori estimation demonstrates that it demands velocity measurements of the satellite with an accuracy of about $10^{-7}$ m/s. Such velocity measurement accuracy is currently difficult to achieve by using existing technology. Therefore, the DCS is used for clock comparison experiments to reduce the requirement for the measurement accuracy of satellite velocity. In most space-ground and space-space clock comparison experiments, the DCS can reduce the effects of the first-order Doppler shift, atmospheric delay, and Shapiro delay. By using the DCS, the Doppler effect is suppressed to the second-order level, greatly relaxing the requirement for the measurement accuracy of satellite velocity. In Eq.(\ref{cl11}), it is assumed that the variation in timing errors of the ground station and the satellite is on the order of 1 ns/s. Usually, the instrument drifts of the signal's reception and transmission do not exceed the variation in timing errors, i.e., $<$1 ns/s. For long-term measurements in the CLEP mission, timing errors and instrumental drifts can be evaluated with high precision. Considering specific space missions, Eq.(\ref{cl11}) can be used to evaluate the test of gravitational redshift.

In order to solve Eqs.(\ref{cl9})-(\ref{cl11}), it is necessary to calculate the influences of individual terms. Since the satellite orbit is the DRO, we use the BCRS to describe the frequency measurement. From the metric of the BCRS \cite{soffel2003iau}, the invariance of Riemann space-time intervals or proper time gives the differential relationship between proper time and coordinate time
\begin{equation}\label{cl12}
 \frac{d\tau_g}{dt} = 1 - \left(\sum_b \frac{U_b(x_g)}{c^2} + \frac{v_{g}^2}{2c^2}\right) + O(c^{-4}),
\end{equation}
\begin{equation}\label{cl13}
 \frac{d\tau_s}{dt} = 1 - \left(\sum_b \frac{U_b(x_s)}{c^2} + \frac{v_{s}^2}{2c^2}\right) + O(c^{-4}),
\end{equation}
where $U_{b}$ represents the gravitational potential of the celestial body $b$ in the Solar System, $v_{g}$ and $v_{s}$ represent the coordinate velocities of the ground station and satellite, respectively. The higher order terms $O(c^{-4})$ are neglected.

Let's consider the terms related to gravitational redshift in the DCS measurement. According to Eqs.(\ref{cl12}) and (\ref{cl13}), the gravitational redshift term in the clock comparison between satellite and ground station is given by
\begin{equation}
\left(\frac{d\tau_s}{dt}\right)_{t_3} \left(\frac{d\tau_g}{dt}\right)_{t_4}^{-1} = 1 - \left(\sum_b \left(\frac{U_b(x_s,t_3)}{c^2} - \frac{U_b(x_g,t_4)}{c^2} \right)+ \frac{v_s^2 }{2c^2}-\frac{v_g^2}{2c^2}\right),
\end{equation}
where the first term in the parentheses represents the gravitational redshift effect, which mainly comes from the gravitational fields of the Earth, Sun, and Moon. For the clock comparison in CLEP, the gravitational redshift due to Earth can reach $6.8\times10^{-10}$, the influence of Earth's oblateness (mainly contributed by the $J_{2}$ term) is less than $4.0\times10^{-16}$, the gravitational redshift of the Sun is no more than $3.1\times10^{-11}$, and the gravitational redshift of the Moon does not exceed $6\times10^{-13}$ (the second column in Table.\ref{table1}). For the tests of the gravitational redshift, clock-comparison experiments measure the deviation between the experimental frequency difference and the predicted value. According to the above estimates, the test of the gravitational redshift mainly depends on the contribution of Earth's mass. The second term in the parentheses represents the time dilation effect or the second-order Doppler effect, which arises from the difference in velocities between the satellite and the ground station. This effect can reach the order of $10^{-10}$. Since the effects of time dilation and gravitational redshift are of the same order, a systematic and accurate evaluation of the time dilation effect is required.
However, it should be noted that in the experiment with Earth-satellite in free fall around the Sun, according to the equivalence principle, the Sun's gravitational redshift cannot be measured and only the tidal effect of the Sun is measured in the clock comparison experiment \cite{PhysRev.121.337,nelson2011relativistic,wolf2016analysis}. Specifically, the Sun's gravitational potential is expressed as $U_{s}(\textbf{\emph{x}}_{a})\cong U_{s}(\textbf{\emph{x}}_{E})+\nabla U_{s}\cdot \textbf{\emph{x}}_{Ea} +U_{tidal}$, and the velocity squared term can be expressed as $\textbf{\emph{v}}_{a}^{2}=\textbf{\emph{v}}^{2}_{E}+2(\frac{d}{dt}(\textbf{\emph{v}}_{E}\cdot\textbf{\emph{x}}_{Ea})-\textbf{\emph{a}}_{E}\cdot\textbf{\emph{x}}_{Ea})+\textbf{\emph{v}}_{Ea}^{2}$, where $\textbf{\emph{x}}_{E}$, $\textbf{\emph{v}}_{E}$, and $\textbf{\emph{a}}_{E}$ are the barycentric position, velocity, and acceleration of the center-of-mass of the Earth, $\textbf{\emph{x}}_{Ea}$ and $\textbf{\emph{v}}_{Ea}$ are the geocentric position and velocity of $a$. Then, the Sun's gravitational redsift is cancelled out by the term associated with $\textbf{\emph{a}}_{E}\cdot\textbf{\emph{x}}_{Ea}$, and only the tidal effect of the Sun can be measured in the clock's measurement. Therefore for CLEP clock comparison, the practical influences of Sun's gravitational field and time dilation effect are much smaller than the priori estimated values directly from the gravitational potential and velocity.

In addition, due to the motion of the ground station in the BCRS, the DCS measurement also includes a variation part of the gravitational redshift term, which can be expressed as
\begin{equation}\label{cld1}
\left(\frac{d\tau_g}{dt}\right)_{t_1} \left(\frac{d\tau_g}{dt}\right)_{t_4}^{-1} = 1 - \left(\sum_b \left(\frac{U_b(x_g,t_1)}{c^2} - \frac{U_b(x_g,t_4)}{c^2}\right) + \frac{v_g^2(t_1)}{2c^2}-\frac{  v_g^2(t_4)}{2c^2}\right).
\end{equation}
The variations of the gravitational redshift and time dilation effects in Eq.(\ref{cld1}) mainly arise from the changes in the position and velocity of the ground station due to the Earth's orbital motion and rotation. The changes in position and velocity can be obtained through Earth's revolution and rotation models. Considering the maximum changes in position and velocity caused by Earth's revolution and rotation, the variation of gravitational redshift is no more than $1.8\times10^{-16}$, and the variation of time dilation effect does not exceed $3\times10^{-14}$. Therefore, for the sake of caution, the contribution of this term needs to be retained in the DCS measurement (note that this can be neglected in close proximity space-ground clock comparisons, such as the ACES mission).

Then, we consider the terms of transmission path of signal, which is related to the coordinate times of the signal's emission and reception. To solve the signal's travel time along the transmission path, the emission time and reception time can be expressed as
\begin{equation}\label{cldr1}
t_3 = t_4 - T_{34} - \Delta t_{34}^{\text{atm}},
\end{equation}
\begin{equation}\label{cldr2}
t_1 = t_4 - T_{34} - T_{12} - \Delta t_{34}^{\text{atm}} - \Delta t_{12}^{\text{atm}} - \Delta t_{23},
\end{equation}
where $T_{12}$ and $T_{34}$ are the time transfer functions between moments $t_{1}$ and $t_{2}$ and $t_{3}$ and $t_{4}$, respectively, which consist of the propagation time in flat spacetime and the gravitational time delay due to gravitational fields. $\Delta t_{12}^{\text{atm}}$ and $\Delta t_{34}^{\text{atm}}$ represent the corresponding atmospheric delays in the transmission paths. $\Delta t_{23}$ represents the signal forwarding delay on the satellite and $\Delta t_{23}$ can be set to be 0 in ideal conditions.

To solve the time transfer function, we set the ground station and satellite velocities and accelerations as $\emph{\textbf{v}}_g(t)$, $\emph{\textbf{a}}_g(t)$, $\emph{\textbf{v}}_s(t)$, and $\emph{\textbf{a}}_s(t)$, respectively. The coordinate distance $R_{34}=|\emph{\textbf{x}}_g(t_4) - \emph{\textbf{x}}_s(t_3)|$ can be expanded by the instantaneous distance $D_{sg}(t_4) = |\emph{\textbf{x}}_g(t_4) - \emph{\textbf{x}}_s(t_4)|$. The time transfer function $T_{34}$ for the downlink can be expressed as
\begin{equation}\label{cldr3}
\begin{split}
T_{34} &= \frac{D_{sg}(t_4)}{c} + \frac{\emph{\textbf{D}}_{sg}(t_4) \cdot \emph{\textbf{v}}_s(t_4)}{c^2}\\
& + \frac{D_{sg}(t_4)}{2c^3} \left[v_s^2(t_4) - \emph{\textbf{D}}_{sg}(t_4) \cdot \emph{\textbf{a}}_s(t_4) + \left(\frac{\emph{\textbf{D}}_{sg}(t_4) \cdot \emph{\textbf{v}}_s(t_4)}{D_{sg}(t_4)}\right)^2\right] + \Delta t_{34}^{gr}.
\end{split}
\end{equation}
The first term represents the instantaneous distance between the satellite and the ground station at time $t_4$. The second and third terms represent the Sagnac effect. The fourth term $\Delta t_{34}^{gr}$ represents the Shapiro delay caused by the gravitational field, which is considered in the subsection.\ref{sus1}. For the time transfer function $T_{12}$ in the uplink, a similar expression can be used
\begin{equation}\label{cldr4}
\begin{split}
T_{12}& = \frac{D_{gs}(t_2)}{c} + \frac{\emph{\textbf{D}}_{gs}(t_2) \cdot \emph{\textbf{v}}_g(t_2)}{c^2} \\
&+ \frac{D_{gs}(t_2)}{2c^3} \left[v_g(t_2)^2 - \emph{\textbf{D}}_{gs}(t_2) \cdot \emph{\textbf{a}}_g(t_2) + \left(\frac{\emph{\textbf{D}}_{gs}(t_2) \cdot \emph{\textbf{v}}_g(t_2)}{D_{gs}(t_2)}\right)^2\right] + \Delta t_{12}^{gr}.
\end{split}
\end{equation}

Combining Eqs.(\ref{cldr1}) and (\ref{cldr3}), the coordinate time ratio $dt_3/dt_4$ in DCS measurement (\ref{cl11}) can be expressed as
\begin{equation}\label{cldr5}
\frac{dt_3}{dt_4}  = 1 - \frac{\emph{\textbf{n}}_{34} \cdot \emph{\textbf{v}}_{34}}{c} - \frac{1}{c^2} \left[(\emph{\textbf{n}}_{34} \cdot \emph{\textbf{v}}_{34})(\emph{\textbf{n}}_{34} \cdot \emph{\textbf{v}}_s(t_4))
+ \emph{\textbf{R}}_{34} \cdot \emph{\textbf{a}}_s(t_4)\right] - \frac{d\Delta t_{34}^{\text{atm}}}{dt_4} - \frac{d\Delta t_{34}^{gr}}{dt_4},
\end{equation}
where $\emph{\textbf{n}}_{34} = \emph{\textbf{R}}_{34} / R_{34}$, $\emph{\textbf{v}}_{34} = \emph{\textbf{v}}_g(t_4) -\emph{\textbf{v}} _s(t_4)$. Note that the Sagnac term of order $c^{-3}$ is ignored in Eq.(\ref{cldr5}) because its effect is canceled out in the DCS measurement. This equation mainly describes the Doppler effect in the one-way downlink. The second term is the first-order Doppler effect and the third term represents the Sagnac effect. The fourth term represents the influences of atmospheric delay, mainly including the influence of tropospheric delay and ionospheric delay. The last term represents the Shapiro delay effect. Clearly, the frequency shift of Eq.(\ref{cldr5}) is dominated by the signal's paths.

Similarly, combining Eqs.(\ref{cldr2})$-$(\ref{cldr4}), another coordinate time ratio $dt_1/dt_4$ in DCS measurement (\ref{cl11}) can be represented as
\begin{equation}\label{cldr6}
\begin{split}
\frac{dt_1}{dt_4} &= 1 - \frac{\emph{\textbf{n}}_{34} \cdot \emph{\textbf{v}}_{34}}{c} - \frac{\emph{\textbf{n}}_{12} \cdot \emph{\textbf{v}}_{12}}{c}- \frac{1}{c^2} \left[(\emph{\textbf{n}}_{34} \cdot \emph{\textbf{v}}_{34})(\emph{\textbf{n}}_{34} \cdot \emph{\textbf{v}}_s(t_4)) + \emph{\textbf{R}}_{34} \cdot \emph{\textbf{a}}_s(t_4)\right] \\
&- \frac{1}{c^2} \left[(\emph{\textbf{n}}_{12} \cdot \emph{\textbf{v}}_{12})(\emph{\textbf{n}}_{12} \cdot \emph{\textbf{v}}_g(t_2)) + \emph{\textbf{R}}_{12} \cdot \emph{\textbf{a}}_g(t_2)\right] \\
&- \frac{d\Delta t_{12}^{gr}}{dt_4} - \frac{d\Delta t_{12}^{\text{atm}}}{dt_4} - \frac{d\Delta t_{34}^{gr}}{dt_4} - \frac{d\Delta t_{34}^{\text{atm}}}{dt_4} - \frac{d\Delta t_{23}}{dt_4},
\end{split}
\end{equation}
where $\emph{\textbf{n}}_{12} = \emph{\textbf{R}}_{12} / R_{12}$, and $\emph{\textbf{v}}_{12} = \emph{\textbf{v}}_s(t_2) - \emph{\textbf{v}}_g(t_1)$. With the same reason as Eq.(\ref{cldr5}), the Sagnac term of order $c^{-3}$ is ignored in Eq.(\ref{cldr6}). This equation $dt_1/dt_4$ mainly describes the Doppler effect in two-way frequency transfer including contributions from the first-order Doppler effect, Sagnac effect, atmospheric shift, and gravitational frequency shift in both uplink and downlink. The last term in Eq.(\ref{cldr6}) represents the influence caused by signal-forwarding delay on the satellite. Under instantaneous forwarding conditions $t_2 = t_3$, the effect of this term is zero. In compensated cases, this term depends on the constant value of compensation, which can be evaluated by the instruments.

Considering the one-way and two-way frequency transfer in DCS measurement, it is the key to combine Eqs.(\ref{cldr5}) and (\ref{cldr6}) to cancel the first-order Doppler shift. Because of the motion of the satellite and ground station, there is the residual Doppler shift in the DCS measurement. From Eqs.(\ref{cldr5}) and (\ref{cldr6}), the residual Doppler shift is given by
\begin{eqnarray}\label{cldr7}
  \frac{dt_3}{dt_4} - \frac{1}{2}\frac{dt_1}{dt_4} &=& \frac{1}{c^2} \left[ (\emph{\textbf{v}}_{43} \cdot \emph{\textbf{v}}_g(t_4)) + \emph{\textbf{R}}_{43} \cdot \emph{\textbf{a}}_g(t_4) \right] \left(1 - \frac{(\emph{\textbf{n}}_{43} \cdot \emph{\textbf{v}}_{43})}{c} \right) \nonumber\\
  &+& \frac{R_{43}}{c^3} \left[ (3\emph{\textbf{v}}_g(t_4) - \emph{\textbf{v}}_s(t_3)) \cdot \emph{\textbf{a}}_g(t_4) - \emph{\textbf{R}}_{43} \cdot \emph{\textbf{d}}_g(t_4) \right] \nonumber\\
   &-& \frac{1}{2}\frac{d(\Delta t_{34}^{gr} - \Delta t_{12}^{gr})}{dt_4} - \frac{1}{2}\frac{d(\Delta t_{34}^{\text{atm}} - \Delta t_{12}^{\text{atm}})}{dt_4} + \frac{d\Delta t_{23}}{dt_4},
\end{eqnarray}
where $\emph{\textbf{d}}_g(t_4) ={d\emph{\textbf{a}}_g(t_4)}/{dt_4}$. The first two terms represent the residual Doppler effect after the Doppler cancellation scheme, where the magnitude of the first term does not exceed $3.5 \times 10^{-10}$ and the second term does not exceed $1.3 \times 10^{-14}$ for the CLEP. The following two terms represent the effects of Shapiro delay, tropospheric delay, and ionospheric delay, which will be evaluated using relevant models and discussed in subsections \ref{sus1}, \ref{sus2}, and \ref{sus3}, respectively. The last term represents the influence of signal forwarding in the satellite, which can be evaluated experimentally.
According to the discussion in subsection \ref{sus1} about Shapiro delay, the gravitational-delay frequency shift can be calculated, and its influence in DCS measurement is much smaller than $1 \times 10^{-17}$. According to the discussions in subsections \ref{sus2} (tropospheric delay) and \ref{sus3} (ionospheric delay), the frequency shift caused by atmospheric delay consists of the frequency shift of the tropospheric delay and ionospheric delay. The tropospheric frequency shift is independent of the transmission signal frequency, while the ionospheric frequency shift depends on the signal frequency. In the CLEP, when considering the same transmission frequency for the uplink and downlink, the total atmospheric delay frequency shift in the DCS measurement is less than $5 \times 10^{-18}$. When the transmission frequencies for the uplink and downlink are not the same, due to the different ionospheric delays between the uplink and downlink, the ionospheric frequency shift in the DCS measurement increases significantly.

The corresponding fractional frequency shifts have been described above for the DCS terms of gravitational redshift, Doppler effect, atmospheric shift, etc. For clock comparison in the CLEP, it is sufficient to ignore terms less than $10^{-16}$.
After some calculations, the DCS measurement (\ref{cl11}) can be reexpressed as
\begin{equation}\label{cldr8}
\begin{split}
\left(\frac{\delta f}{f}\right)_{\text{dcs}}
&= \left( \sum_{b} \frac{{U_{b}(\emph{\textbf{x}}_{s},t_3) - U_{b}(\emph{\textbf{x}}_{g},t_4)}}{{c^2}} + \frac{{v_s^2(t_3) - v_g^2(t_4)}}{{2c^2}} \right) \left( 1 - \frac{{ \emph{\textbf{n}}_{34} \cdot\emph{\textbf{v}}_{34}}}{{c}} \right) \\
&\quad -  \frac{{1}}{{2}}\left(  \sum_{b}\frac{{U_{b}(\emph{\textbf{x}}_{g},t_1) - U_{b}(\emph{\textbf{x}}_{g},t_4)}}{{c^2}}+\frac{{v_g^2(t_1) - v_g^2(t_4)}}{{2c^2}}\right)\\
&\quad+ \left(\frac{{ \emph{\textbf{v}}_{43} \cdot  \emph{\textbf{v}}_{g}(t_4) +  \emph{\textbf{R}}_{43} \cdot  \emph{\textbf{a}}_{g}(t_4)}}{{c^{2}}}\right) \left( 1- \frac{{ \emph{\textbf{n}}_{34} \cdot\emph{\textbf{v}}_{34}}}{{c}}\right)\\
& \quad+ \frac{R_{43}}{c^3}\left[ \left(3 \emph{\textbf{v}}_{g}(t_4) -  \emph{\textbf{v}}_{s}(t_3) \right)\cdot  \emph{\textbf{a}}_{g}(t_{4}) +  \emph{\textbf{R}}_{43} \cdot  \emph{\textbf{d}}_{g}(t_4)\right] \\
&\quad-\frac{d(\Delta t_{34}^{gr} - \Delta t_{12}^{gr})}{2dt_2} -\frac{ d(\Delta t_{34}^{\text{atm}} - \Delta t_{12}^{\text{atm}})}{2 dt_4}
+\frac{{d\Delta t_{23}}}{{dt_4}} \\
&\quad + \left(  \frac{ d\delta \tau_{s}^t}{dt_3} +\frac{ d\delta \tau_{s}^e}{dt_3}  - \frac{d\delta \tau_{g}^t}{dt_4} - \frac{d\delta \tau_{g}^r}{dt_4} \right) \left( 1 - \frac{{ \emph{\textbf{n}}_{34} \cdot\emph{\textbf{v}}_{34}}}{{c}} \right)\\
& \quad-\frac{{1}}{{2}} \left(\frac{ d\delta \tau_{g}^e}{dt_1}  - \frac{d\delta \tau_{g}^r}{dt_4}\right) \left( 1 - 2\frac{{ \emph{\textbf{n}}_{34} \cdot\emph{\textbf{v}}_{34}}}{{c}} \right).
\end{split}
\end{equation}
This equation is dominated by gravitational redshift and time dilation and describes the gravitational redshift measurement of the CLEP's clock comparison with DCS. It also includes the Doppler effect, tropospheric frequency shift, ionospheric frequency shift, and gravitational frequency shift of the DCS measurement. Based on the basic parameter of CLEP mission, the estimations of various effects can be found in the third column of Table \ref{table1}.

The measurement error of positions and velocity will lead to the corresponding error of frequency measurements. Taking into account the position error and velocity error of the ground station, as well as the position error and velocity error of the satellite, the error in frequency measurement with DCS can be calculated as
\begin{equation}\label{cldr9}
\begin{split}
u(\delta f/f)_{\text{dcs}}& = \bigg{[} \left(\frac{{\partial(\delta f/f)_{\text{dcs}}}}{{\partial \emph{\textbf{x}}_g}} \cdot \delta\emph{\textbf{x}}_g\right)^2 + \left(\frac{{\partial(\delta f/f)_{\text{dcs}}}}{{\partial \emph{\textbf{v}}_g}} \cdot \delta \emph{\textbf{v}}_g\right)^2\nonumber\\
&+ \left(\frac{{\partial(\delta f/f)_{\text{dcs}}}}{{\partial \emph{\textbf{x}}_s}} \cdot \delta \emph{\textbf{x}}_s\right)^2 + \left(\frac{{\partial(\delta f/f)_{\text{dcs}}}}{{\partial \emph{\textbf{v}}_s}} \cdot \delta \emph{\textbf{v}}_s\right)^2 \bigg{]}^{1/2},
\end{split}
\end{equation}
where $\delta\emph{\textbf{x}}_g$ and $\delta\emph{\textbf{v}}_g$ represent the position error and velocity error of the ground station, $\delta\emph{\textbf{x}}_s$ and $\delta\emph{\textbf{v}}_s$ are the position error and velocity error of the satellite, respectively.
Considering a position accuracy of 1 m for the ground station relative to the Earth's center, a position accuracy of 10 km for the satellite orbit, and a position accuracy of 1 km for the Earth within the Solar System, a priori estimation demonstrates that the error in gravitational redshift for the Earth is $6.6\times10^{-16}$, for the Sun is $6.4\times10^{-16}$, and for the Moon is $5.4\times10^{-17}$ (the fourth column as shown in Table \ref{table1}).
Considering a velocity accuracy of 0.1 m/s for the satellite and the center-of-mass of the Earth, an error in time dilation effect is $<1.6\times10^{-15}$, an error in the residual Doppler effect is $<3\times10^{-15}$, an error in atmospheric frequency shift is $<1\times10^{-18}$, and an error in gravitational-delay frequency shift is $<1\times10^{-18}$ (the fourth column as shown in Table \ref{table1}).

\begin{table}[!t]
\centering
\caption{\label{table1} The magnitude estimation of various effects for one-day CLEP simulated experiment. The first column represents the types of effects in clock comparison. The second column indicates the magnitude of various effects in the one-way measurement, the third column is the magnitude of various effects in the DCS measurement, and the fourth column is the estimated error of various effects in the DCS measurement. }
\newcommand{\tabincell}[2]{\begin{tabular}{@{}#1@{}}#2\end{tabular}}
\begin{tabular}{lccc}
\hline
\tabincell{l}{Type of effect}      &Magnitude            &DCS measurement     &Error \\
\hline
First-order Doppler effect         &$10^{-6}\sim10^{-5}$       & $<1\times10^{-10}$       &$<3\times10^{-15}$ \\
Earth's gravitational redshift     &$\sim6.9\times10^{-10}$    &$\sim6.9\times10^{-10}$   &$<6.6\times10^{-16}$ \\
Sun's gravitational redshift       &$\sim3\times10^{-11}$      &$\sim3\times10^{-11}$       &$<6.4\times10^{-16}$ \\
Moon's gravitational redshift      &$\sim6\times10^{-13}$      &$\sim6\times10^{-13}$      &$<5\times10^{-17}$ \\
Time dilation effect               &$<6\times10^{-10}$       &$<2\times10^{-11}$       &$<1.6\times10^{-15}$ \\
Tropospheric frequency shift       &$\sim10^{-12}$             &$<1\times10^{-16}$      &$<1\times10^{-17}$ \\
Ionospheric frequency shift        &$\sim10^{-13}$          & $<1\times10^{-15}$        &$<2\times10^{-16}$ \\
Gravitational frequency shift      &$\sim10^{-14}$           & $<1\times10^{-17}$         &$<1\times10^{-18}$ \\
Temperature effect                 &$-$           & $-$         &$<1\times10^{-15}$ \\
\hline
\end{tabular}
\end{table}

\subsection{Gravitational frequency shift}\label{sus1}
From Einstein's general relativity, the gravitational field can lead to an extra time delay in light travel time, known as Shapiro delay. Assuming that the metric is given, the standard methods can be obtained by solving the null geodesic equation or the eikonal equation \cite{misner1973gravitation,PhysRevD.94.124007,ashby2010accurate}. Furthermore, different approaches are developed to study the light propagation in the gravitational fields based on the Synge World function and time transfer function \cite{le2004world,teyssandier2008general,PhysRevD.89.064045,PhysRevD.93.044028}. Based on these methods, many solutions have been proposed in the post-Newtonian and post-Minkowskian approximations \cite{zschocke2010efficient,PhysRevD.86.044007,PhysRevD.90.084020,PhysRevD.96.024003,PhysRevD.97.024045,PhysRevD.100.064063}. In the post-Newtonian approximation, the metric can be expanded as $g^{\alpha\beta}=\eta^{\alpha\beta}+h^{\alpha\beta}$, where $\eta_{\alpha\beta}=$diag($-$1,+1,+1,+1), and $h_{\alpha\beta}$ represents the gravitational perturbation. In the gravitational field of the solar system, the Shapiro delay of light propagation can be calculated by
\begin{equation}\label{gtd1}
\Delta t_{34}^{gr} = \frac{R_{34}}{2c} \int_0^1 \left(h^{00}-2n_{34}^i h^{0i}+n_{34}^in_{34}^j h^{ij}\right)_{x_{(l)}} dl,
\end{equation}
where $n_{34}^i = ({\emph{\textbf{x}} _4-\emph{\textbf{x}}_3})^{i}/{R_{34}}$. For this equation, the integral is calculated along the straight line from the satellite to the ground station that is defined by the parameter equation $\emph{\textbf{x}}(l) = l(\emph{\textbf{x}}_4-\emph{\textbf{x}}_3) + \emph{\textbf{x}}_3$ with $0\leq l \leq1$, where $l$ is an affine parameter. The gravitational time delay $\Delta t_{12}^{gr}$ in the uplink is given by a similar method. For a crude estimation, the Shapiro delay caused by the Earth's mass corresponds to the level of $10^{-10}$ seconds.

From Eq.(\ref{cldr8}), the gravitational frequency shift is given by the time derivative of the Shapiro delay, such as the gravitational frequency shift in the downlink ${d\Delta t_{34}^{gr}}/{dt_4}$. After some calculations, the gravitational frequency shift is given by
\begin{equation}\label{gtd2}
\begin{split}
\left(\frac{\delta f}{f}\right)^{gr}_{34} = & -\sum_b \frac{GM_b(r_{b3}+r_{b4})}{c^3 r_{b3} r_{b4}} \bigg{[} \left(\frac{2}{1+\emph{\textbf{n}}_{b3}\cdot \emph{\textbf{n}}_{b4}}-\frac{r_{b3}-r_{b4}}{r_{b3}+r_{b4}} \right) \emph{\textbf{n}}_{34}\cdot(\emph{\textbf{v}}_{b3}-\emph{\textbf{v}}_{b4}) \\
&+ \frac{2R_{34}}{r_{b3}+r_{b4}}\frac{\emph{\textbf{n}}_{b3}\cdot \emph{\textbf{v}}_{b3}+\emph{\textbf{n}}_{b4}\cdot \emph{\textbf{v}}_{b4}}{1+\emph{\textbf{n}}_{b3}\cdot \emph{\textbf{n}}_{b4}}\bigg{]},
\end{split}
\end{equation}
where subscripts $\emph{\textbf{n}}_{b3}$ and $\emph{\textbf{n}}_{b4}$ represent the unit vectors pointing from the center-of-mass of celestial body $b$ to the satellite at time $t_{3}$ and to ground station at time $t_4$, respectively. $\emph{\textbf{v}}_{b3}$ and $\emph{\textbf{v}}_{b4}$ represent the velocities of the satellite at time $t_3$ and ground station at time $t_4$ relative to the celestial body $b$, respectively. For a crude estimation, the Sun's gravitational frequency shift is less than $7\times10^{-14}$. Considering the contribution from the Earth, the effect of this term is less than $6\times10^{-15}$. In the case of the Moon, it is much smaller than $1\times10^{-16}$ and can be safely ignored. The gravitational frequency shift in the uplink path $(\delta f/f)^{gr}_{12}$ can be expressed in the same form by replacing 3 with 1 and 4 with 2. With the DCS, the contributions of the gravitational frequency shift in the one-way and two-way paths canceled out and the residual effect is no greater than $1\times10^{-17}$ with the error smaller than $1\times10^{-18}$ (as shown in Table.\ref{table1}).

\subsection{Tropospheric frequency shift}\label{sus2}
For the influence of the atmosphere, it can be split into tropospheric delay and ionospheric delay $\Delta t^{\text{atm}} = \Delta t^{\text{tro}} + \Delta t^{\text{ion}}$.
The tropospheric delay is related to the pressure, temperature, and humidity along the signal transmission path, which can be represented as \cite{PhysRevD.108.064031,shen2016formulation}
\begin{equation}\label{tro1}
\Delta t^{\text{tro}} = \frac{1}{c} \int \left({M_1+M_2} \right) dl_{t},
\end{equation}
where $l_{t}$ represents the signal's path in the troposphere, $M_1 = 77.6 \left({p}/{T}\right) \times 10^{-6} {K}/{\text{mbar}}$ and $M_2 = 0.373 \left({\epsilon}/{T^2}\right) {K^2}/{\text{mbar}}$, $T$ represents temperature, $p$ represents total pressure, and $\epsilon$ represents the water vapor partial pressure along the signal's path $l_{t}$.
Therefore, the frequency shift due to the tropospheric delay can be expressed as
\begin{equation}
\left(\frac{\delta f}{f}\right)^{\text{tro}}_{34} =\frac{d\Delta t_{34}^{\text{tro}}}{dt_4} = \frac{1}{c} \frac{d}{dt_4} \int \left({M_1+M_2}\right) dl_{t34},
\end{equation}
where $l_{t34}$ represents the tropospheric path of the signal in the downlink. From this equation, a one-way fractional frequency shift due to tropospheric delay is related to the variations of parameters $M_1$ and $M_2$. It can be concluded that the changes in temperature, total pressure, water vapor partial pressure, and signal's geometric path will lead to the tropospheric frequency shift. The tropospheric delay $\Delta t_{12}^{\text{tro}}$ can be expressed in a similar form. Since the uplink and downlink paths are close, the differences in parameters $M_1$ and $M_2$ between the uplink and downlink paths can be ignored. Therefore, the frequency shift due to tropospheric delay is dominated by the geometric path variation of the signal. Under the above approximation, the frequency shift of the tropospheric delay in DCS measurement can be expressed as
\begin{equation}\label{tr01}
\left(\frac{\delta f}{f}\right)^{\text{tro}}_{\text{dcs}}= {\frac{d\Delta t_{12}^{\text{tro}}}{2dt_4} - \frac{d\Delta t_{34}^{\text{tro}}}{2dt_4}} = \frac{\overline{M}_1+\overline{M}_2}{c} (|V_{t12}| - |V_{t34}|),
\end{equation}
where $\overline{M}_1$ and $\overline{M}_2$ represent the average values of parameters $M_1$ and $M_2$ along the tropospheric transmission paths $l_{t12}$ and $l_{t34}$, respectively, and $|V_{t12}|$ and $|V_{t34}|$ represent the velocities of the tropospheric transmission path variations $l_{t12}$ and $l_{t34}$, respectively. For the influences of the tropospheric delay or the ionospheric delay, the fractional frequency shifts depend on the changes in the total path delay and are related to the velocity between the satellite and the ground station. Considering above equations, a prior estimate indicates that the tropospheric frequency shift is no greater than the level of $8 \times 10^{-13}$ in the one-way frequency transfer, is no greater than the level of $1\times10^{-16}$ in the DCS measurement, and it's error is less than the level of $1\times10^{-17}$ (as shown in Table. \ref{table1}).

\subsection{Ionospheric frequency shift}\label{sus3}

The ionospheric delay is dependent on the signal frequency $f$ and is primarily dominated by the $f^{-2}$ term. It can be expressed as $\Delta t^{\text{ion}} = -\left({40.3}/{cf^2}\right) \int n_e dl_{i}$, where $n_{e}$ represents the electron density along the path. Therefore, in the downlink, the frequency shift due to ionospheric delay can be written as \cite{PhysRevD.108.064031,shen2016formulation}
\begin{equation}\label{3232}
  \left(\frac{\delta f}{f}\right)^{\text{ion}}_{34} =\frac{d\Delta t_{34}^{\text{ion}}}{dt_4} = -\frac{40.3}{cf^2} \frac{d}{dt_4} \int n_e dl_{i34},
\end{equation}
where $l_{i34}$ represents the ionospheric path of the signal in the downlink. In the one-way frequency shift, s simple estimation shows that the fractional frequency shift of ionospheric delay can reach the level of $10^{-13}$. For the uplink, the fractional frequency shift of ionospheric delay $\left({\delta f}/{f}\right)^{\text{tro}}_{12}$ can be represented in the same form. Similar to the tropospheric case, since the uplink and downlink paths are very close, the differences in electron density along the uplink and downlink paths can be ignored. Considering that two downlink frequencies are different, the fractional frequency shift of ionospheric delay in the DCS measurement can be expressed as
\begin{equation}\label{io01}
\left(\frac{\delta f}{f}\right)^{\text{ion}}_{\text{dcs}}= \left(\frac{d\Delta t_{12}^{\text{ion}}}{2dt_4} - \frac{d\Delta t_{34}^{\text{ion}}}{2dt_4}\right)= \frac{40.3 \overline{n}_{e}}{2c}\left(\frac{|V_{i12}|}{f^2_{2w}} - \frac{|V_{i34}|}{f^{2}_{1w}}\right),
\end{equation}
where $\overline{n}_{e}$ is the average electron density along the ionospheric path of the signal, $f_{1w}$ is the frequency of one-way signal, $f_{2w}$ is the frequency of two-way signal, and $|V_{i12}|$ and $|V_{i34}|$ represent the velocities of the ionospheric transmission path variations $l_{i12}$ and $l_{i34}$, respectively.
Assuming the downlink frequency in one-way signal is 24 GHz and the downlink frequency in one-way signal is 25 GHz, a prior estimate demonstrates that the ionospheric frequency shift is no greater than the level of $2\times 10^{-14}$ in the one-way signal, is no greater than the level of $1\times10^{-15}$ in the DCS measurement, and it's error is less than the level of $2\times10^{-16}$ with 20 $\%$ error of ionospheric electron density model \cite{NAVA20081856} (as shown in Table. \ref{table1}). By adopting closer frequencies of one-way and two-way signals, the ionospheric influence can be further suppressed.

\subsection{Clock error}
\begin{figure}
  \centering
  \includegraphics[width=0.6\textwidth]{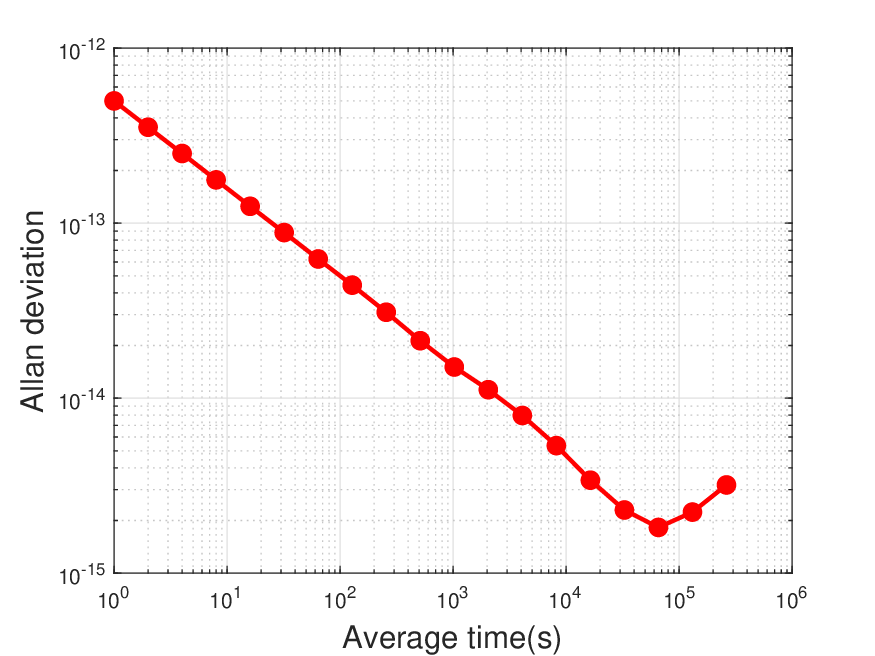}
  \caption{The simulated Allan deviation of onboard passive hydrogen maser as a function of the average time. The stability is about $5\times 10^{-13}/\sqrt{\tau}$.  }\label{fig2}
\end{figure}

In the CLEP mission, we consider that satellite clock is a passive hydrogen maser and the ground-station clock is a H clock of equal or better performance. The frequency of hydrogen maser is perturbed by a superposition of several types of independent noise. Considering the stochastic noise nature of the clocks, the one-sided power spectra density of fractional frequency fluctuations can be expressed as \cite{barnes1971characterization,lesage1979characterization}
\begin{equation}\label{ceor1}
  S(f)=\sum_{n=-2}^{2}h_{n}f^{n}
\end{equation}
where $h_{i}$ is the noise intensity coefficient, and $f$ is the Fourier frequency variable. There are five classic types of noise: the white phase modulation noise ($f^{2}$), the flicker phase modulation noise ($f^1$), the white frequency modulation noise ($f^{0}$), the flicker frequency modulation noise ($f^{-1}$), and the random walk frequency modulation noise ($f^{-2}$).
Based on the five types of clock noise, We can generate a series of clock frequency errors.
For the onboard passive hydrogen maser, we set the stability as $5\times 10^{-13}/\sqrt{\tau}$ ($\tau$ is the averaging time in seconds). The simulated Allan deviation of onboard passive hydrogen maser is showed in the figure \ref{fig2} (the stability and accuracy of ground-station hydrogen clock are better, which is not shown here). The simulation result shows that the stability of onboard clock is about $5\times 10^{-13}/\sqrt{\tau}$ and reaches the magnitude of $2\times 10^{-15}$ after 60 000 s.
For hydrogen masers, there is a long-term frequency drift, which is shown with the curve in Fig. \ref{fig2} after 100 000 s. For long-term clock comparisons, the influences of the frequency drift of hydrogen clocks will be significant, which can affect the test of gravitational redshift. In addition, the frequency offset of onboard H-maser also has an impact on the test of gravitational redshift. For high-precision experiments to test gravitational redshift, the accurate measurement of this drift is a huge challenge. For this challenge, a potential method may consider multiple ground stations to establish frequency links with satellite for clock comparisons. The frequency offset and drift of the on-board clock are the same in the clock comparison for different ground stations, which may be used to consider the influences of frequency offset and drift when fitting the data. The same fractional frequency shift in different data sets may help to assess or reduce the influences of frequency offset and drift of the on-board clock. Another possible solution is for CLEP satellites to carry a clock with much better absolute accuracy, such as a cold atom clock.
For simplicity, in our simulated experiment, we assume that frequency drift and offset can be accurately measured without affecting the simulated evaluation of gravitational redshift, although it is very difficult in the experiment.

\subsection{Other influences}
In addition to the factors discussed above, there are many other effects that can affect satellite-ground clock comparison and the test of gravitational redshift, such as solar radiation pressure, temperature changes, magnetic field variations, etc. Detailed discussions on these effects of various factors are complex and is beyond this work. Here we only briefly discuss the influences of solar radiation pressure, temperature, and magnetic fields.

The solar radiation pressure can have various effects on the CLEP clock comparison experiment, such as satellite's  orbit and attitude. We mainly consider the impact on the change of orbit position and velocity, assuming that other effects are small and can be handled with the corresponding technologies.
Considering the influence of solar radiation pressure, the accurate calculation can be given by a numerical model based on the spacecraft's surface geometry and optical characteristics of its components. In the case considered here, it is sufficient to use a simplified analytical model known as the ``box-wing" model. We employ the ``box-wing" model to calculate the influence of solar radiation pressure, denoted as $\delta \emph{\textbf{x}}_{sr}$. Solar radiation pressure arises from the absorption and emission of photons on the surface exposed to sunlight. To describe radiation pressure, we consider the surface area of the satellite as $A_{\text{plane}}$, and use unit vectors $\mathbf{e}_n$ and $\mathbf{e}_{\bigodot}$ to represent the surface normal direction and the direction of the light source, respectively. For this given plane, the acceleration caused by solar radiation pressure can be expressed as \cite{rodriguez2012adjustable}
\begin{equation}\label{srp}
\emph{\textbf{a}}_{sr} = -\frac{\Phi A_{\text{plane}} \cos\theta}{mc}\left[(\eta+\delta)\left(\mathbf{e}_{\bigodot}+\frac{2}{3}\mathbf{e}_n\right) + 2\rho\cos\theta\mathbf{e}_n\right],
\end{equation}
where $m$ is the total mass of the satellite, $\Phi = \left({{\left(1\, \text{AU}\right)}^2}/{{r_{\bigodot}}^2}\right)$ 1367 W/m$^2$ represents the radiation flux per unit area near the Earth, $1 \text{AU} = 1.496\times 10^8$ km, denotes the average distance between the Earth and the Sun, $r_{\bigodot}$ is the instantaneous distance from the satellite to the Sun, $\theta$ is the angle between unit vectors $\mathbf{e}_n$ and $\mathbf{e}_{\bigodot}$, and $\eta$, $\delta$, and $\rho$ respectively denote the absorption rate, diffuse reflectance, and specular reflectance of the satellite's surface with relationship $\eta+\delta+\rho=1$. In Eq.(\ref{srp}), the term in parentheses can be regarded as a weighted average $\mathbf{e}_{\bigodot}+({2}/{3})\mathbf{e}_n$ of the solar direction and the surface normal direction $\mathbf{e}_n$. Considering the relationship among absorption rate, diffuse reflectance, and specular reflectance, it can be concluded that the influence of solar radiation pressure mainly depends on the specular reflectance parameter $\rho$. From a crude estimation, the acceleration caused by solar radiation pressure is on the order of 100 nm/s$^2$, and the resulting position changes $\delta \emph{\textbf{x}}_{sr}$ is smaller than the measurement error of the satellite's position 10 km.

Regarding the influence of temperature variations, it is dominated by the Sun's radiation. We conservatively estimate the influence of the temperature by considering the temperature sensitivity of the clock and the maximum variations of the temperature at the clock's location with respect to the orientation of the satellite. Similarly, the sensitivity of the onboard clock to temperature is assumed to be the same as that of a ground laboratory clock. The temperature sensitivity of a ground laboratory hydrogen clock is characterized in fractional frequency $<2\times 10^{-14}$ K$^{-1}$ \cite{rochat2012atomic}. The variation of the temperature can be modeled by the satellite's and Sun's orientations. Moreover, the clock is controlled actively by the thermal pipes and the influence of temperature can be evaluated. The temperature monitoring accuracy is less than 0.05 K in ideal condition, and the effect of temperature on the clock can be assessed to the level better than $1\times10^{-15}$.

Regarding the influence of magnetic fields, the International Geomagnetic Reference Field (IGRF-12) model can be used to estimated the magnetic effects on the on-board clocks of low-orbit satellites \cite{thebault2015evaluation}. For orbital altitudes above 1 000 km, the magnetic field model is complicated due to the influence of the solar wind.
It includes the magnetopause, which points toward the Sun, and the magnetotail, which points away from the Sun. The CLEP satellite is in a higher orbit than the magnetopause, but may be in the magnetotail.
In the case that the onboard clock adopts magnetic shielding technology, the influence of magnetic fields can be ignored.
In the case of no magnetic shielding technology and a large magnetic field effect, one may discard the data in the magnetotail region.

\section{Estimation of test of gravitational redshift}\label{stgr}

To evaluate the accuracy of the gravitational redshift test in the CLEP experiment, we use the DCS model for simulated experiments. We consider that the clocks on the ground station and in satellite are H clocks, their fractional frequency instability and fractional inaccuracy are $5\times 10^{-13}/\sqrt{\tau}$, where $\tau$ is the measurement averaging time in seconds. The experiments compare the frequencies between the ground clock and satellite clock through a microwave link. In short measurement times, the microwave link noise is the main noise. In long measurement times, the measurement noise is mainly dominated by clock frequency noise. The experiment enables long measurement time or continuous clock comparisons. The gravitational field models of the Moon and Sun are considered point-particle models and the gravitational field model of the Earth is based on EGM2008 \cite{pavlis2012development}.

\begin{figure}
  \centering
  \includegraphics[width=0.6\textwidth]{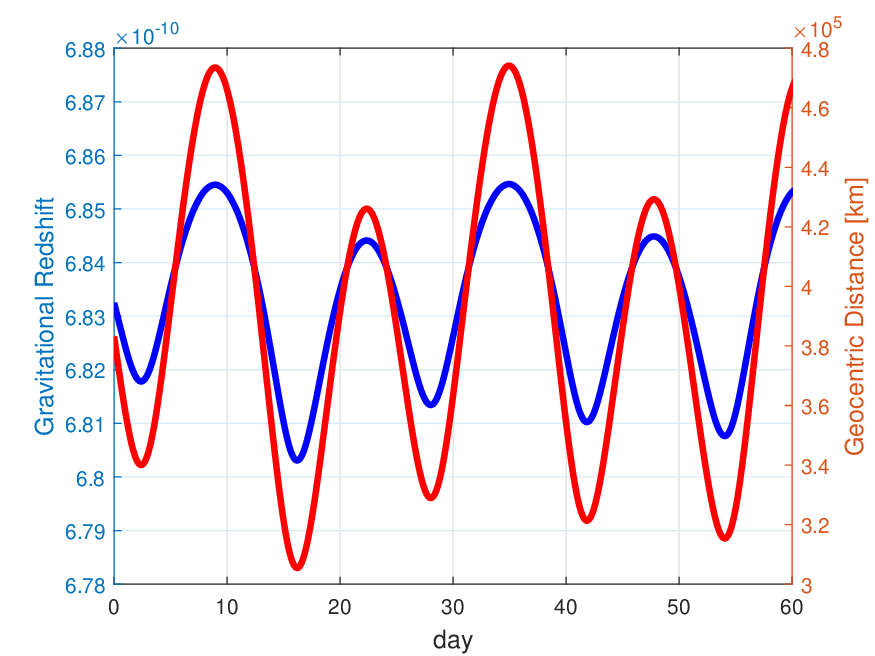}
  \caption{The predicted value of gravitational redshift in the CLEP experiment. The blue line represents the gravitational redshift and the red line indicates the geocentric distance of the CLEP satellite around the Moon. }\label{fig3}
\end{figure}

In the test of gravitational redshift, the DCS measurement includes gravitational redshift, time dilation effect, and other residual effects. These effects in the DCS measurement are simulated by the simulated positions and velocities of satellite and ground station clock. From the simulated DCS measurements, we estimate the accuracy of testing gravitational redshift. To perform the analysis of frequency comparison data, we define a quantity $Y(t)$ that is defined by the simulated data minus the theoretical value of relativistic model, which is used to obtain the violation parameter $\alpha$. Considering the space-ground clock comparison at a given time $t_{i}$, the quantity $Y(t_i)$ is described as
\begin{equation}\label{gres1}
  Y(t_{i})= \Delta y(t_i)-\sum_{b}\frac{\Delta U_{b}(t_i)}{c^2}-\frac{\Delta \emph{\textbf{v}}^{2}(t_i)}{2c^2}-\delta_{\text{ot}},
\end{equation}
where $\Delta y(t_i)$ is the simulated DCS measurement value of frequency comparison, the second term is predicted value of gravitational redshift, the third term is predicted value of time dilation effect, and $\delta_{\text{ot}}$ represents the predicted value of other residual effects in Eq.(\ref{cldr8}).

Based on the simulation and model of DCS measurement, we can use the least-squares estimator for the determination of the parameter $\alpha$. With the matrix form, we define a fitting model with equation of length $N$
\begin{equation}\label{fitm1}
  Y=X\beta + \epsilon,
\end{equation}
where $\beta$ is the parameter vector to be estimated that is used to estimate $\alpha$, $\epsilon$ is the noise vector, which is assumed gaussian with value of expectation $E[\epsilon]=0$ and standard deviation $\sigma$, $Y$ is given by
\begin{equation}\label{fitm2}
  Y=\left[ Y(t_1)\,\,\,Y(t_2)\,\,\,...\,\,\,Y(t_N)\right]^{\text{T}},
\end{equation}
where $Y(t_{i})$ is the quantity given by Eq.(\ref{gres1}) at coordinate time $t_{i}$, and $X$ is given by
\begin{equation}\label{fitm3}
  X=\left[ \Delta y_{gr}(t_1)\,\,\,\Delta y_{gr}(t_2)\,\,\,...\,\,\,\Delta y_{gr}(t_N)\right]^{\text{T}},
\end{equation}
where $\Delta y_{gr}(t_i)=\sum_{b}{\Delta U_{b}(t_i)}/{c^2}$ represents the theoretical prediction value of the standard gravitational redshift at coordinate time $t_{i}$.

\begin{figure}
  \centering
  \includegraphics[width=0.6\textwidth]{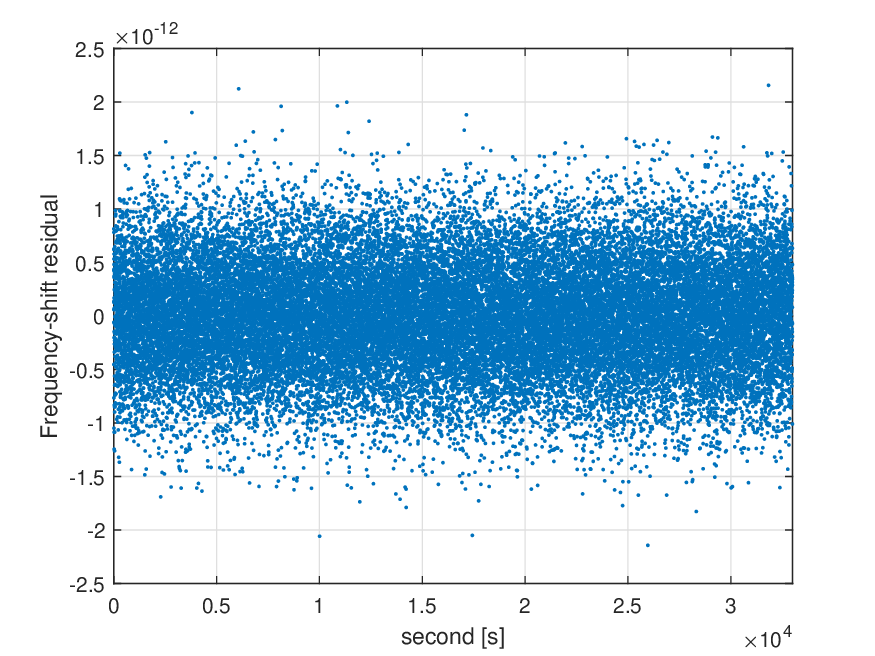}
  \caption{The simulated residual frequency shift of the DCS measurement. The blue dots is given by subtracting the predicted DCS frequency shift from the simulated DCS data. }\label{fig4}
\end{figure}

Considering the clock comparison between the CLEP satellite and ground station, we choose the ground station as Beijing Miyun station (the position is given by $116.976^{\circ}$ E, $40.368^{\circ}$ N, 160.0 m altitude). We investigate the convergence of the EGM2008 by considering the evolution of the estimated uncertainty of gravitational redshift on the ground station, which is given by the difference of the $(n-1)$-order potential and $n$-order potential of the Earth. When the difference in gravitational redshift is smaller than $1\times10^{-16}$, we can neglect the higher terms of the EGM2008, and the 20-order Earth gravitational model is sufficient for the CLEP experiment (since the satellite clock is in a high orbit, the effect on the satellite clock is different from the ground clock, but it is smaller). Base on the 20-order model of EGM2008, Figure.\ref{fig3} shows the simulated gravitational redshift effect for 60 days, and its magnitude can reach about $6.9\times 10^{-10}$. For a conservative estimation, we set the observation cutoff elevation angle of the CLEP satellite larger than $30^{\circ}$ on the ground station, which results in an average of several hours of observations per day. Figure.\ref{fig4} shows the residual frequency shift of a one-day experiment that is given by subtracting the theoretical predicted DCS frequency shift from the simulated DCS data and the residual is dominated by the clock noise. With the least-squares method and simulated DCS measurement, we obtain a simulated result for the potential test of gravitational redshift  $(0.4\pm3.7)\times10^{-6}$ for 60-day observations, where $0.4\times10^{-6}$ represent mean offset between the theoretical true value and simulation result and $\pm3.7\times10^{-6}$ is the uncertainty. The relatively small value of $\alpha$ is not surprising since the model assumes $\alpha = 0$ and the simulation should therefore give a value of $\alpha$ well within the uncertainty. The important result is the uncertainty. After a simulated measurement session of 60 days under ideal conditions, the CLEP experiment may reach the uncertainty of $3.7\times10^{-6}$ for potential test of gravitational redshift.

\section{Conclusion}\label{concl}
General relativity, as one of the foundational theories of modern physics, provides the most precise description of the macroscopic physical world and our universe. The accurate measurement of general relativity's validity is essential for advancing our understanding of fundamental physics. Gravitational redshift, a classic test of general relativity, can be tested with high precision using space-based missions. China's Lunar Exploration Project (CLEP) includes the proposals of using an onboard passive hydrogen clock with daily stability of $2\times 10^{-15}$. By establishing a high-precision frequency link between the ground station and the satellite's hydrogen clock, CLEP has a potential to perform precise tests of gravitational redshift.

Based on the CLEP experiment, we developed clock-comparison models for one-way frequency measurement, two-way frequency measurement, and DCS measurement, respectively. Considering the clock comparison between the CLEP satellite and the ground station, we analyzed the influences of various factors, such as the Doppler effect, satellite orbit determination, atmospheric delay, Shapiro delay, etc. The Doppler cancellation scheme measurement model effectively reduces the first-order Doppler effect, significantly relaxing the requirement on velocity measurement accuracy in gravitational redshift measurements.

Furthermore, we conducted simulations to compare the clocks between the CLEP satellite and the ground station and evaluated the empirical precision of the gravitational redshift effect. Our simulation results demonstrate that under ideal condition of high-precision measurement of the onboard H-maser frequency offset and drift, the CLEP experiment has a potential to achieve an uncertainty of 3-4 ppm in testing gravitational redshift after a 60-day measurement session, and a longer measurement duration could potentially yield even better results. For such test with deep-space missions, some engineering and technical challenges still require significant effort to overcome, such as high-precision measurements of onboard-clock frequency offset and drift.

\section{Acknowledgment}
The authors thank the anonymous reviewers for their comments and constructive suggestions that greatly improve this work. This work is supported by the National Natural Science Foundation of China (Grants No.12247150, No.12305062, No.12175076 and No.11925503), the Post-doctoral Science Foundation of China (Grant No.2022M721257), and Strategic Priority Research Program on Space Science, the Chinese Academy of Sciences (XDA30040400).

\section{References}
\bibliographystyle{elsarticle-num}
\bibliography{redgr}

\begin{thebibliography}{10}
\expandafter\ifx\csname url\endcsname\relax
  \def\url#1{\texttt{#1}}\fi
\expandafter\ifx\csname urlprefix\endcsname\relax\def\urlprefix{URL }\fi
\expandafter\ifx\csname href\endcsname\relax
  \def\href#1#2{#2} \def\path#1{#1}\fi

\bibitem{will2014confrontation}
C.~M. Will, The confrontation between general relativity and experiment, Living
  reviews in relativity 17 (2014) 1--117.

\bibitem{taylor1988dilaton}
T.~Taylor, G.~Veneziano, Dilaton couplings at large distances, Physics Letters
  B 213~(4) (1988) 450--454.

\bibitem{damour1994string}
T.~Damour, A.~M. Polyakov, The string dilation and a least coupling principle,
  Nuclear Physics B 423~(2-3) (1994) 532--558.

\bibitem{rubakov2001large}
V.~A. Rubakov, Large and infinite extra dimensions, Physics-Uspekhi 44~(9)
  (2001) 871.

\bibitem{maartens2010brane}
R.~Maartens, K.~Koyama, Brane-world gravity, Living Reviews in Relativity 13
  (2010) 1--124.

\bibitem{wagner2012torsion}
T.~A. Wagner, S.~Schlamminger, J.~Gundlach, E.~G. Adelberger, Torsion-balance
  tests of the weak equivalence principle, Classical and Quantum Gravity
  29~(18) (2012) 184002.

\bibitem{PhysRevLett.100.041101}
S.~Schlamminger, K.-Y. Choi, T.~A. Wagner, J.~H. Gundlach, E.~G. Adelberger,
  \href{https://link.aps.org/doi/10.1103/PhysRevLett.100.041101}{Test of the
  equivalence principle using a rotating torsion balance}, Phys. Rev. Lett. 100
  (2008) 041101.
\newblock \href {https://doi.org/10.1103/PhysRevLett.100.041101}
  {\path{doi:10.1103/PhysRevLett.100.041101}}.
\newline\urlprefix\url{https://link.aps.org/doi/10.1103/PhysRevLett.100.041101}

\bibitem{PhysRevLett.129.121102}
P.~Touboul, G.~M\'etris, M.~Rodrigues, J.~Berg\'e, A.~Robert, Q.~Baghi,
  Y.~Andr\'e, J.~Bedouet, D.~Boulanger, S.~Bremer, P.~Carle, R.~Chhun,
  B.~Christophe, V.~Cipolla, T.~Damour, P.~Danto, L.~Demange, H.~Dittus,
  O.~Dhuicque, P.~Fayet, B.~Foulon, P.-Y. Guidotti, D.~Hagedorn, E.~Hardy,
  P.-A. Huynh, P.~Kayser, S.~Lala, C.~L\"ammerzahl, V.~Lebat, F.~m.~c. Liorzou,
  M.~List, F.~L\"offler, I.~Panet, M.~Pernot-Borr\`as, L.~Perraud, S.~Pires,
  B.~Pouilloux, P.~Prieur, A.~Rebray, S.~Reynaud, B.~Rievers, H.~Selig,
  L.~Serron, T.~Sumner, N.~Tanguy, P.~Torresi, P.~Visser,
  \href{https://link.aps.org/doi/10.1103/PhysRevLett.129.121102}{$microscope$
  mission: Final results of the test of the equivalence principle}, Phys. Rev.
  Lett. 129 (2022) 121102.
\newblock \href {https://doi.org/10.1103/PhysRevLett.129.121102}
  {\path{doi:10.1103/PhysRevLett.129.121102}}.
\newline\urlprefix\url{https://link.aps.org/doi/10.1103/PhysRevLett.129.121102}

\bibitem{PhysRevLett.123.231102}
H.~P.-l. Bars, C.~Guerlin, A.~Hees, R.~Peaucelle, J.~D. Tasson, Q.~G. Bailey,
  G.~Mo, P.~Delva, F.~Meynadier, P.~Touboul, G.~M\'etris, M.~Rodrigues,
  J.~Berg\'e, P.~Wolf,
  \href{https://link.aps.org/doi/10.1103/PhysRevLett.123.231102}{New test of
  lorentz invariance using the microscope space mission}, Phys. Rev. Lett. 123
  (2019) 231102.
\newblock \href {https://doi.org/10.1103/PhysRevLett.123.231102}
  {\path{doi:10.1103/PhysRevLett.123.231102}}.
\newline\urlprefix\url{https://link.aps.org/doi/10.1103/PhysRevLett.123.231102}

\bibitem{PhysRevLett.118.221102}
P.~Delva, J.~Lodewyck, S.~Bilicki, E.~Bookjans, G.~Vallet, R.~Le~Targat, P.-E.
  Pottie, C.~Guerlin, F.~Meynadier, C.~Le~Poncin-Lafitte, O.~Lopez,
  A.~Amy-Klein, W.-K. Lee, N.~Quintin, C.~Lisdat, A.~Al-Masoudi, S.~D\"orscher,
  C.~Grebing, G.~Grosche, A.~Kuhl, S.~Raupach, U.~Sterr, I.~R. Hill, R.~Hobson,
  W.~Bowden, J.~Kronj\"ager, G.~Marra, A.~Rolland, F.~N. Baynes, H.~S.
  Margolis, P.~Gill,
  \href{https://link.aps.org/doi/10.1103/PhysRevLett.118.221102}{Test of
  special relativity using a fiber network of optical clocks}, Phys. Rev. Lett.
  118 (2017) 221102.
\newblock \href {https://doi.org/10.1103/PhysRevLett.118.221102}
  {\path{doi:10.1103/PhysRevLett.118.221102}}.
\newline\urlprefix\url{https://link.aps.org/doi/10.1103/PhysRevLett.118.221102}

\bibitem{qin2023testing}
C.-G. Qin, J.~Ke, Q.~Li, Y.-F. Chen, J.~Luo, Y.-J. Tan, C.-G. Shao, Testing
  lorentz symmetry with space-based gravitational-wave detectors, Classical and
  Quantum Gravity (2023).

\bibitem{sanner2019optical}
C.~Sanner, N.~Huntemann, R.~Lange, C.~Tamm, E.~Peik, M.~S. Safronova, S.~G.
  Porsev, Optical clock comparison for lorentz symmetry testing, Nature
  567~(7747) (2019) 204--208.

\bibitem{PhysRevLett.109.080801}
J.~Gu\'ena, M.~Abgrall, D.~Rovera, P.~Rosenbusch, M.~E. Tobar, P.~Laurent,
  A.~Clairon, S.~Bize,
  \href{https://link.aps.org/doi/10.1103/PhysRevLett.109.080801}{Improved tests
  of local position invariance using $^{87}\mathrm{Rb}$ and $^{133}\mathrm{Cs}$
  fountains}, Phys. Rev. Lett. 109 (2012) 080801.
\newblock \href {https://doi.org/10.1103/PhysRevLett.109.080801}
  {\path{doi:10.1103/PhysRevLett.109.080801}}.
\newline\urlprefix\url{https://link.aps.org/doi/10.1103/PhysRevLett.109.080801}

\bibitem{PhysRevA.87.010102}
S.~Peil, S.~Crane, J.~L. Hanssen, T.~B. Swanson, C.~R. Ekstrom,
  \href{https://link.aps.org/doi/10.1103/PhysRevA.87.010102}{Tests of local
  position invariance using continuously running atomic clocks}, Phys. Rev. A
  87 (2013) 010102.
\newblock \href {https://doi.org/10.1103/PhysRevA.87.010102}
  {\path{doi:10.1103/PhysRevA.87.010102}}.
\newline\urlprefix\url{https://link.aps.org/doi/10.1103/PhysRevA.87.010102}

\bibitem{ashby2018null}
N.~Ashby, T.~E. Parker, B.~R. Patla, A null test of general relativity based on
  a long-term comparison of atomic transition frequencies, Nature Physics
  14~(8) (2018) 822--826.

\bibitem{PhysRevLett.126.011102}
R.~Lange, N.~Huntemann, J.~M. Rahm, C.~Sanner, H.~Shao, B.~Lipphardt, C.~Tamm,
  S.~Weyers, E.~Peik,
  \href{https://link.aps.org/doi/10.1103/PhysRevLett.126.011102}{Improved
  limits for violations of local position invariance from atomic clock
  comparisons}, Phys. Rev. Lett. 126 (2021) 011102.
\newblock \href {https://doi.org/10.1103/PhysRevLett.126.011102}
  {\path{doi:10.1103/PhysRevLett.126.011102}}.
\newline\urlprefix\url{https://link.aps.org/doi/10.1103/PhysRevLett.126.011102}

\bibitem{PhysRevLett.3.439}
R.~V. Pound, G.~A. Rebka,
  \href{https://link.aps.org/doi/10.1103/PhysRevLett.3.439}{Gravitational
  red-shift in nuclear resonance}, Phys. Rev. Lett. 3 (1959) 439--441.
\newblock \href {https://doi.org/10.1103/PhysRevLett.3.439}
  {\path{doi:10.1103/PhysRevLett.3.439}}.
\newline\urlprefix\url{https://link.aps.org/doi/10.1103/PhysRevLett.3.439}

\bibitem{PhysRevLett.4.337}
R.~V. Pound, G.~A. Rebka,
  \href{https://link.aps.org/doi/10.1103/PhysRevLett.4.337}{Apparent weight of
  photons}, Phys. Rev. Lett. 4 (1960) 337--341.
\newblock \href {https://doi.org/10.1103/PhysRevLett.4.337}
  {\path{doi:10.1103/PhysRevLett.4.337}}.
\newline\urlprefix\url{https://link.aps.org/doi/10.1103/PhysRevLett.4.337}

\bibitem{PhysRev.140.B788}
R.~V. Pound, J.~L. Snider,
  \href{https://link.aps.org/doi/10.1103/PhysRev.140.B788}{Effect of gravity on
  gamma radiation}, Phys. Rev. 140 (1965) B788--B803.
\newblock \href {https://doi.org/10.1103/PhysRev.140.B788}
  {\path{doi:10.1103/PhysRev.140.B788}}.
\newline\urlprefix\url{https://link.aps.org/doi/10.1103/PhysRev.140.B788}

\bibitem{PhysRevLett.45.2081}
R.~F.~C. Vessot, M.~W. Levine, E.~M. Mattison, E.~L. Blomberg, T.~E. Hoffman,
  G.~U. Nystrom, B.~F. Farrel, R.~Decher, P.~B. Eby, C.~R. Baugher, J.~W.
  Watts, D.~L. Teuber, F.~D. Wills,
  \href{https://link.aps.org/doi/10.1103/PhysRevLett.45.2081}{Test of
  relativistic gravitation with a space-borne hydrogen maser}, Phys. Rev. Lett.
  45 (1980) 2081--2084.
\newblock \href {https://doi.org/10.1103/PhysRevLett.45.2081}
  {\path{doi:10.1103/PhysRevLett.45.2081}}.
\newline\urlprefix\url{https://link.aps.org/doi/10.1103/PhysRevLett.45.2081}

\bibitem{PhysRevLett.121.231101}
P.~Delva, N.~Puchades, E.~Sch\"onemann, F.~Dilssner, C.~Courde, S.~Bertone,
  F.~Gonzalez, A.~Hees, C.~Le~Poncin-Lafitte, F.~Meynadier, R.~Prieto-Cerdeira,
  B.~Sohet, J.~Ventura-Traveset, P.~Wolf,
  \href{https://link.aps.org/doi/10.1103/PhysRevLett.121.231101}{Gravitational
  redshift test using eccentric galileo satellites}, Phys. Rev. Lett. 121
  (2018) 231101.
\newblock \href {https://doi.org/10.1103/PhysRevLett.121.231101}
  {\path{doi:10.1103/PhysRevLett.121.231101}}.
\newline\urlprefix\url{https://link.aps.org/doi/10.1103/PhysRevLett.121.231101}

\bibitem{PhysRevLett.121.231102}
S.~Herrmann, F.~Finke, M.~L\"ulf, O.~Kichakova, D.~Puetzfeld, D.~Knickmann,
  M.~List, B.~Rievers, G.~Giorgi, C.~G\"unther, H.~Dittus, R.~Prieto-Cerdeira,
  F.~Dilssner, F.~Gonzalez, E.~Sch\"onemann, J.~Ventura-Traveset,
  C.~L\"ammerzahl,
  \href{https://link.aps.org/doi/10.1103/PhysRevLett.121.231102}{Test of the
  gravitational redshift with galileo satellites in an eccentric orbit}, Phys.
  Rev. Lett. 121 (2018) 231102.
\newblock \href {https://doi.org/10.1103/PhysRevLett.121.231102}
  {\path{doi:10.1103/PhysRevLett.121.231102}}.
\newline\urlprefix\url{https://link.aps.org/doi/10.1103/PhysRevLett.121.231102}

\bibitem{takamoto2020test}
M.~Takamoto, I.~Ushijima, N.~Ohmae, T.~Yahagi, K.~Kokado, H.~Shinkai,
  H.~Katori, Test of general relativity by a pair of transportable optical
  lattice clocks, Nature Photonics 14~(7) (2020) 411--415.

\bibitem{takano2016geopotential}
T.~Takano, M.~Takamoto, I.~Ushijima, N.~Ohmae, T.~Akatsuka, A.~Yamaguchi,
  Y.~Kuroishi, H.~Munekane, B.~Miyahara, H.~Katori, Geopotential measurements
  with synchronously linked optical lattice clocks, Nature Photonics 10~(10)
  (2016) 662--666.

\bibitem{grotti2018geodesy}
J.~Grotti, S.~Koller, S.~Vogt, S.~H{\"a}fner, U.~Sterr, C.~Lisdat, H.~Denker,
  C.~Voigt, L.~Timmen, A.~Rolland, et~al., Geodesy and metrology with a
  transportable optical clock, Nature Physics 14~(5) (2018) 437--441.

\bibitem{bothwell2022resolving}
T.~Bothwell, C.~J. Kennedy, A.~Aeppli, D.~Kedar, J.~M. Robinson, E.~Oelker,
  A.~Staron, J.~Ye, Resolving the gravitational redshift across a
  millimetre-scale atomic sample, Nature 602~(7897) (2022) 420--424.

\bibitem{zheng2023lab}
X.~Zheng, J.~Dolde, M.~C. Cambria, H.~M. Lim, S.~Kolkowitz, A lab-based test of
  the gravitational redshift with a miniature clock network, Nature
  Communications 14~(1) (2023) 4886.

\bibitem{zheng2022differential}
X.~Zheng, J.~Dolde, V.~Lochab, B.~N. Merriman, H.~Li, S.~Kolkowitz,
  Differential clock comparisons with a multiplexed optical lattice clock,
  Nature 602~(7897) (2022) 425--430.

\bibitem{hinkley2013atomic}
N.~Hinkley, J.~A. Sherman, N.~B. Phillips, M.~Schioppo, N.~D. Lemke, K.~Beloy,
  M.~Pizzocaro, C.~W. Oates, A.~D. Ludlow, An atomic clock with 10--18
  instability, Science 341~(6151) (2013) 1215--1218.

\bibitem{bloom2014optical}
B.~Bloom, T.~Nicholson, J.~Williams, S.~Campbell, M.~Bishof, X.~Zhang,
  W.~Zhang, S.~Bromley, J.~Ye, An optical lattice clock with accuracy and
  stability at the 10- 18 level, Nature 506~(7486) (2014) 71--75.

\bibitem{PhysRevLett.116.063001}
N.~Huntemann, C.~Sanner, B.~Lipphardt, C.~Tamm, E.~Peik,
  \href{https://link.aps.org/doi/10.1103/PhysRevLett.116.063001}{Single-ion
  atomic clock with
  $3\ifmmode\times\else\texttimes\fi{}{10}^{\ensuremath{-}18}$ systematic
  uncertainty}, Phys. Rev. Lett. 116 (2016) 063001.
\newblock \href {https://doi.org/10.1103/PhysRevLett.116.063001}
  {\path{doi:10.1103/PhysRevLett.116.063001}}.
\newline\urlprefix\url{https://link.aps.org/doi/10.1103/PhysRevLett.116.063001}

\bibitem{mcgrew2018atomic}
W.~McGrew, X.~Zhang, R.~Fasano, S.~Sch{\"a}ffer, K.~Beloy, D.~Nicolodi,
  R.~Brown, N.~Hinkley, G.~Milani, M.~Schioppo, et~al., Atomic clock
  performance enabling geodesy below the centimetre level, Nature 564~(7734)
  (2018) 87--90.

\bibitem{PhysRevLett.123.033201}
S.~M. Brewer, J.-S. Chen, A.~M. Hankin, E.~R. Clements, C.~W. Chou, D.~J.
  Wineland, D.~B. Hume, D.~R. Leibrandt,
  \href{https://link.aps.org/doi/10.1103/PhysRevLett.123.033201}{$^{27}{\mathrm{al}}^{+}$
  quantum-logic clock with a systematic uncertainty below
  ${10}^{\ensuremath{-}18}$}, Phys. Rev. Lett. 123 (2019) 033201.
\newblock \href {https://doi.org/10.1103/PhysRevLett.123.033201}
  {\path{doi:10.1103/PhysRevLett.123.033201}}.
\newline\urlprefix\url{https://link.aps.org/doi/10.1103/PhysRevLett.123.033201}

\bibitem{PhysRevApplied.17.034041}
Y.~Huang, B.~Zhang, M.~Zeng, Y.~Hao, Z.~Ma, H.~Zhang, H.~Guan, Z.~Chen,
  M.~Wang, K.~Gao,
  \href{https://link.aps.org/doi/10.1103/PhysRevApplied.17.034041}{Liquid-nitrogen-cooled
  $\mathrm{Ca}$${}^{+}$ optical clock with systematic uncertainty of
  $3\ifmmode\times\else\texttimes\fi{}{10}^{\ensuremath{-}18}$}, Phys. Rev.
  Appl. 17 (2022) 034041.
\newblock \href {https://doi.org/10.1103/PhysRevApplied.17.034041}
  {\path{doi:10.1103/PhysRevApplied.17.034041}}.
\newline\urlprefix\url{https://link.aps.org/doi/10.1103/PhysRevApplied.17.034041}

\bibitem{meynadier2018atomic}
F.~Meynadier, P.~Delva, C.~le~Poncin-Lafitte, C.~Guerlin, P.~Wolf, Atomic clock
  ensemble in space (aces) data analysis, Classical and Quantum Gravity 35~(3)
  (2018) 035018.

\bibitem{savalle2019gravitational}
E.~Savalle, C.~Guerlin, P.~Delva, F.~Meynadier, C.~le~Poncin-Lafitte, P.~Wolf,
  Gravitational redshift test with the future aces mission, Classical and
  Quantum Gravity 36~(24) (2019) 245004.

\bibitem{bongs2015development}
K.~Bongs, Y.~Singh, L.~Smith, W.~He, O.~Kock, D.~{\'S}wierad, J.~Hughes,
  S.~Schiller, S.~Alighanbari, S.~Origlia, et~al., Development of a strontium
  optical lattice clock for the soc mission on the iss, Comptes Rendus Physique
  16~(5) (2015) 553--564.

\bibitem{derevianko2022fundamental}
A.~Derevianko, K.~Gibble, L.~Hollberg, N.~R. Newbury, C.~Oates, M.~S.
  Safronova, L.~C. Sinclair, N.~Yu, Fundamental physics with a state-of-the-art
  optical clock in space, Quantum Science and Technology 7~(4) (2022) 044002.

\bibitem{PhysRevD.108.064031}
W.~Shen, P.~Zhang, Z.~Shen, R.~Xu, X.~Sun, M.~Ashry, A.~Ruby, W.~Xu, K.~Wu,
  Y.~Wu, A.~Ning, L.~Wang, L.~Li, C.~Cai,
  \href{https://link.aps.org/doi/10.1103/PhysRevD.108.064031}{Testing
  gravitational redshift based on microwave frequency links onboard the china
  space station}, Phys. Rev. D 108 (2023) 064031.
\newblock \href {https://doi.org/10.1103/PhysRevD.108.064031}
  {\path{doi:10.1103/PhysRevD.108.064031}}.
\newline\urlprefix\url{https://link.aps.org/doi/10.1103/PhysRevD.108.064031}

\bibitem{PhysRevD.107.064032}
F.~De~Marchi, G.~Cascioli, T.~Ely, L.~Iess, E.~A. Burt, S.~Hensley,
  E.~Mazarico,
  \href{https://link.aps.org/doi/10.1103/PhysRevD.107.064032}{Testing the
  gravitational redshift with an inner solar system probe: The veritas case},
  Phys. Rev. D 107 (2023) 064032.
\newblock \href {https://doi.org/10.1103/PhysRevD.107.064032}
  {\path{doi:10.1103/PhysRevD.107.064032}}.
\newline\urlprefix\url{https://link.aps.org/doi/10.1103/PhysRevD.107.064032}

\bibitem{nunes2023gravitational}
N.~Nunes, N.~Bartel, A.~Belonenko, G.~Manucharyan, S.~Popov, V.~Rudenko,
  L.~Gurvits, G.~Cim{\`o}, G.~M. Calv{\'e}s, M.~Zakhvatkin, et~al.,
  Gravitational redshift test of eep with radioastron from near earth to the
  distance of the moon, Classical and Quantum Gravity 40~(17) (2023) 175005.

\bibitem{zheng2018technical}
W.~Zheng, J.~Zhang, G.~Wang, R.~Zhu, L.~Tong, F.~Tong, F.~Shu, L.~Liu, Z.~Sun,
  Technical progress of the chinese vlbi network, Channels 16~(120) (2018) 16.

\bibitem{zhi2022experiment}
W.~Zhi-Chao, L.~Qing-Hui, Z.~Xin, Z.~Juan, X.~Yong-Hui, D.~Tao, J.~Jian-Hua,
  Z.~Chao, W.~Ling-Ling, L.~Yue, The experiment and analysis on vlbi
  observations of space passive hydrogen maser, Chinese Astronomy and
  Astrophysics 46~(3) (2022) 297--308.

\bibitem{blanchet2001relativistic}
L.~Blanchet, C.~Salomon, P.~Teyssandier, P.~Wolf, Relativistic theory for time
  and frequency transfer to order, Astronomy \& Astrophysics 370~(1) (2001)
  320--329.

\bibitem{PhysRevD.66.024045}
B.~Linet, P.~Teyssandier,
  \href{https://link.aps.org/doi/10.1103/PhysRevD.66.024045}{Time transfer and
  frequency shift to the order ${1/c}^{4}$ in the field of an axisymmetric
  rotating body}, Phys. Rev. D 66 (2002) 024045.
\newblock \href {https://doi.org/10.1103/PhysRevD.66.024045}
  {\path{doi:10.1103/PhysRevD.66.024045}}.
\newline\urlprefix\url{https://link.aps.org/doi/10.1103/PhysRevD.66.024045}

\bibitem{qin2019relativistic}
C.-G. Qin, Y.-J. Tan, C.-G. Shao, Relativistic tidal effects on
  clock-comparison experiments, Classical and Quantum Gravity 36~(5) (2019)
  055008.

\bibitem{jia2023investigation}
Q.~Jia, Q.~Li, J.~Liang, L.~Liu, Investigation of proper time and
  inter-satellite clock difference using general relativity theory, Aerospace
  Science and Technology 132 (2023) 108071.

\bibitem{soffel2003iau}
M.~Soffel, S.~A. Klioner, G.~Petit, P.~Wolf, S.~Kopeikin, P.~Bretagnon,
  V.~Brumberg, N.~Capitaine, T.~Damour, T.~Fukushima, et~al., The iau 2000
  resolutions for astrometry, celestial mechanics, and metrology in the
  relativistic framework: explanatory supplement, The Astronomical Journal
  126~(6) (2003) 2687.

\bibitem{PhysRev.121.337}
B.~Hoffmann,
  \href{https://link.aps.org/doi/10.1103/PhysRev.121.337}{Noon-midnight red
  shift}, Phys. Rev. 121 (1961) 337--342.
\newblock \href {https://doi.org/10.1103/PhysRev.121.337}
  {\path{doi:10.1103/PhysRev.121.337}}.
\newline\urlprefix\url{https://link.aps.org/doi/10.1103/PhysRev.121.337}

\bibitem{nelson2011relativistic}
R.~A. Nelson, Relativistic time transfer in the vicinity of the earth and in
  the solar system, Metrologia 48~(4) (2011) S171--S180.

\bibitem{wolf2016analysis}
P.~Wolf, L.~Blanchet, Analysis of sun/moon gravitational redshift tests with
  the ste-quest space mission, Classical and Quantum Gravity 33~(3) (2016)
  035012.

\bibitem{misner1973gravitation}
C.~W. Misner, K.~S. Thorne, J.~A. Wheeler, Gravitation, Macmillan, 1973.

\bibitem{PhysRevD.94.124007}
S.~Zschocke, \href{https://link.aps.org/doi/10.1103/PhysRevD.94.124007}{Light
  propagation in the gravitational field of one arbitrarily moving pointlike
  body in the 2pn approximation}, Phys. Rev. D 94 (2016) 124007.
\newblock \href {https://doi.org/10.1103/PhysRevD.94.124007}
  {\path{doi:10.1103/PhysRevD.94.124007}}.
\newline\urlprefix\url{https://link.aps.org/doi/10.1103/PhysRevD.94.124007}

\bibitem{ashby2010accurate}
N.~Ashby, B.~Bertotti, Accurate light-time correction due to a gravitating
  mass, Classical and Quantum Gravity 27~(14) (2010) 145013.

\bibitem{le2004world}
C.~Le~Poncin-Lafitte, B.~Linet, P.~Teyssandier, World function and time
  transfer: general post-minkowskian expansions, Classical and Quantum Gravity
  21~(18) (2004) 4463.

\bibitem{teyssandier2008general}
P.~Teyssandier, C.~Le~Poncin-Lafitte, General post-minkowskian expansion of
  time transfer functions, Classical and Quantum Gravity 25~(14) (2008) 145020.

\bibitem{PhysRevD.89.064045}
A.~Hees, S.~Bertone, C.~Le~Poncin-Lafitte,
  \href{https://link.aps.org/doi/10.1103/PhysRevD.89.064045}{Relativistic
  formulation of coordinate light time, doppler, and astrometric observables up
  to the second post-minkowskian order}, Phys. Rev. D 89 (2014) 064045.
\newblock \href {https://doi.org/10.1103/PhysRevD.89.064045}
  {\path{doi:10.1103/PhysRevD.89.064045}}.
\newline\urlprefix\url{https://link.aps.org/doi/10.1103/PhysRevD.89.064045}

\bibitem{PhysRevD.93.044028}
B.~Linet, P.~Teyssandier,
  \href{https://link.aps.org/doi/10.1103/PhysRevD.93.044028}{Time transfer
  functions in schwarzschild-like metrics in the weak-field limit: A unified
  description of shapiro and lensing effects}, Phys. Rev. D 93 (2016) 044028.
\newblock \href {https://doi.org/10.1103/PhysRevD.93.044028}
  {\path{doi:10.1103/PhysRevD.93.044028}}.
\newline\urlprefix\url{https://link.aps.org/doi/10.1103/PhysRevD.93.044028}

\bibitem{zschocke2010efficient}
S.~Zschocke, S.~A. Klioner, On the efficient computation of the quadrupole
  light deflection, Classical and Quantum Gravity 28~(1) (2010) 015009.

\bibitem{PhysRevD.86.044007}
X.-M. Deng, Y.~Xie,
  \href{https://link.aps.org/doi/10.1103/PhysRevD.86.044007}{Two-post-newtonian
  light propagation in the scalar-tensor theory: An $n$-point mass case}, Phys.
  Rev. D 86 (2012) 044007.
\newblock \href {https://doi.org/10.1103/PhysRevD.86.044007}
  {\path{doi:10.1103/PhysRevD.86.044007}}.
\newline\urlprefix\url{https://link.aps.org/doi/10.1103/PhysRevD.86.044007}

\bibitem{PhysRevD.90.084020}
A.~Hees, S.~Bertone, C.~Le~Poncin-Lafitte,
  \href{https://link.aps.org/doi/10.1103/PhysRevD.90.084020}{Light propagation
  in the field of a moving axisymmetric body: Theory and applications to the
  juno mission}, Phys. Rev. D 90 (2014) 084020.
\newblock \href {https://doi.org/10.1103/PhysRevD.90.084020}
  {\path{doi:10.1103/PhysRevD.90.084020}}.
\newline\urlprefix\url{https://link.aps.org/doi/10.1103/PhysRevD.90.084020}

\bibitem{PhysRevD.96.024003}
C.-G. Qin, C.-G. Shao,
  \href{https://link.aps.org/doi/10.1103/PhysRevD.96.024003}{General
  post-minkowskian expansion and application of the phase function}, Phys. Rev.
  D 96 (2017) 024003.
\newblock \href {https://doi.org/10.1103/PhysRevD.96.024003}
  {\path{doi:10.1103/PhysRevD.96.024003}}.
\newline\urlprefix\url{https://link.aps.org/doi/10.1103/PhysRevD.96.024003}

\bibitem{PhysRevD.97.024045}
C.~Jiang, W.~Lin,
  \href{https://link.aps.org/doi/10.1103/PhysRevD.97.024045}{Post-newtonian
  light propagation in kerr-newman spacetime}, Phys. Rev. D 97 (2018) 024045.
\newblock \href {https://doi.org/10.1103/PhysRevD.97.024045}
  {\path{doi:10.1103/PhysRevD.97.024045}}.
\newline\urlprefix\url{https://link.aps.org/doi/10.1103/PhysRevD.97.024045}

\bibitem{PhysRevD.100.064063}
C.-G. Qin, Y.-J. Tan, Y.-F. Chen, C.-G. Shao,
  \href{https://link.aps.org/doi/10.1103/PhysRevD.100.064063}{Light propagation
  in the field of the $n$-body system and its application in the tianqin
  mission}, Phys. Rev. D 100 (2019) 064063.
\newblock \href {https://doi.org/10.1103/PhysRevD.100.064063}
  {\path{doi:10.1103/PhysRevD.100.064063}}.
\newline\urlprefix\url{https://link.aps.org/doi/10.1103/PhysRevD.100.064063}

\bibitem{shen2016formulation}
Z.~Shen, W.-B. Shen, S.~Zhang, Formulation of geopotential difference
  determination using optical-atomic clocks onboard satellites and on ground
  based on doppler cancellation system, Geophysical Journal International
  206~(2) (2016) 1162--1168.

\bibitem{NAVA20081856}
B.~Nava, P.~Coïsson, S.~Radicella,
  \href{https://www.sciencedirect.com/science/article/pii/S1364682608000357}{A
  new version of the nequick ionosphere electron density model}, Journal of
  Atmospheric and Solar-Terrestrial Physics 70~(15) (2008) 1856--1862,
  ionospheric Effects and Telecommunications.
\newblock \href {https://doi.org/https://doi.org/10.1016/j.jastp.2008.01.015}
  {\path{doi:https://doi.org/10.1016/j.jastp.2008.01.015}}.
\newline\urlprefix\url{https://www.sciencedirect.com/science/article/pii/S1364682608000357}

\bibitem{barnes1971characterization}
J.~A. Barnes, A.~R. Chi, L.~S. Cutler, D.~J. Healey, D.~B. Leeson, T.~E.
  McGunigal, J.~A. Mullen, W.~L. Smith, R.~L. Sydnor, R.~F. Vessot, et~al.,
  Characterization of frequency stability, IEEE transactions on instrumentation
  and measurement~(2) (1971) 105--120.

\bibitem{lesage1979characterization}
P.~Lesage, C.~Audoin, Characterization and measurement of time and frequency
  stability, Radio Science 14~(4) (1979) 521--539.

\bibitem{rodriguez2012adjustable}
C.~Rodriguez-Solano, U.~Hugentobler, P.~Steigenberger, Adjustable box-wing
  model for solar radiation pressure impacting gps satellites, Advances in
  Space Research 49~(7) (2012) 1113--1128.

\bibitem{rochat2012atomic}
P.~Rochat, F.~Droz, Q.~Wang, S.~Froidevaux, Atomic clocks and timing systems in
  global navigation satellite systems, in: Proceedings of European navigation
  conference, Gdansk, April, 2012, pp. 1--11.

\bibitem{thebault2015evaluation}
E.~Th{\'e}bault, C.~C. Finlay, P.~Alken, C.~D. Beggan, E.~Canet, A.~Chulliat,
  B.~Langlais, V.~Lesur, F.~J. Lowes, C.~Manoj, et~al., Evaluation of candidate
  geomagnetic field models for igrf-12, Earth, Planets and Space 67~(1) (2015)
  1--23.

\bibitem{pavlis2012development}
N.~K. Pavlis, S.~A. Holmes, S.~C. Kenyon, J.~K. Factor, The development and
  evaluation of the earth gravitational model 2008 (egm2008), Journal of
  geophysical research: solid earth 117~(B4) (2012).

\end{thebibliography}

\end{document}